\newlength\smallfigwidth
\newlength\figwidth
\documentclass[aps,floatfix,amsmath,amssymb,showkeys,letterpaper,twocolumn]{revtex4}

\newcommand{\be}{\begin{equation}}
\newcommand{\ee}{\end{equation}}
\newcommand{\bn}{\begin{eqnarray}}
\newcommand{\en}{\end{eqnarray}}
\newcommand{\ii}{{\rm i}}
\newcommand{\cn}{{\rm c}_{\bf n}}
\newcommand{\cm}{{\rm c}_{\bf m}}
\newcommand{\sn}{{\rm s}_{\bf n}}

\usepackage{graphicx}
\usepackage{amsmath}
\usepackage{bm}

\usepackage{color}
\definecolor{blue}{rgb}{0.0,0.0,0.4}
\definecolor{red}{rgb}{0.5,0.0,0.0}
\definecolor{purple}{rgb}{0.3,0.0,0.6}
\definecolor{black}{rgb}{0.0,0.0,0.0}
\newcommand{\change}[1]{\textcolor{black}{#1}}

\begin{document}

\title{Metastability and dynamics in remanent states of square artificial spin ice
with long-range dipole interactions}

  \author{G.\ M.\  Wysin}
  \email{wysin@phys.ksu.edu}
  \homepage{http://www.phys.ksu.edu/personal/wysin}
  \affiliation{Department of Physics, Kansas State University, Manhattan, KS 66506-2601}

\date{18 October 2023} 
\begin{abstract}
After removal of an applied magnetic field, artificial square spin ice can be left in a
metastable remanent state, with nonzero residual magnetization and excess energy above the ground state.
Using a model of magnetic islands with dipoles of fixed magnitude and local anisotropies, 
the remanent states are precisely determined here, including all long-range dipole interactions.
Small deviations away from remanent states are analyzed and the frequencies of modes of oscillation are determined.
Some modes reach zero frequency at high symmetry wave vectors, such that the stability limits are found,
as determined by the local anisotropy strength relative to the dipolar coupling strength.
\end{abstract}
\maketitle

\section{Remanent states in square artificial spin ice}
\label{intro}

\begin{figure}
\includegraphics[width=1.2\smallfigwidth,angle=0]{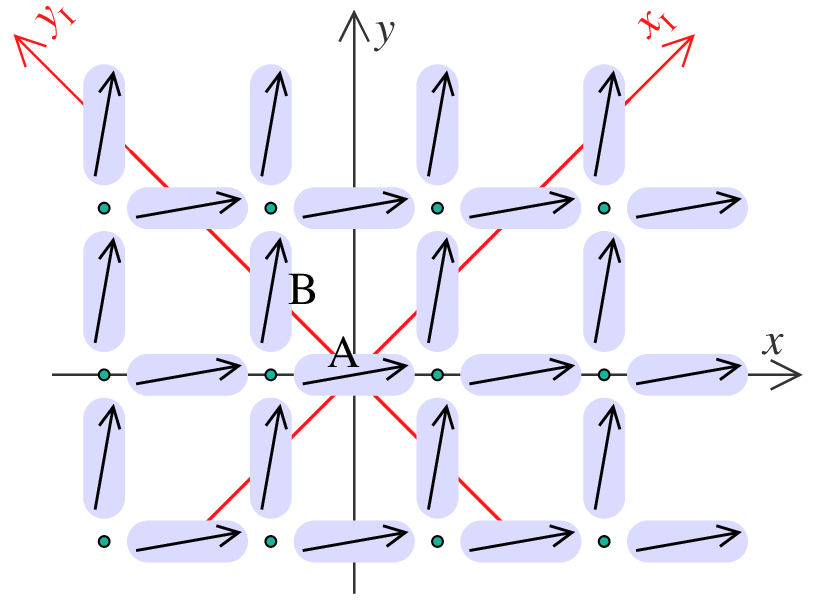}
\caption{\label{square-rs} Square spin ice in a remanent state magnetized along \change{a principal} {\sc nn}-direction $x_{\rm I}$. 
Dots indicate the vertices, each surrounded by four islands. {\bf A} and {\bf B} represent the two
sublattices for the two island orientations.  The island lattice constant along diagonal directions is 
$a_{\rm I}=a_{\rm v}/\sqrt{2}$, where $a_{\rm v}$ is the vertex lattice constant along \change{the original $xy$
coordinates of the square lattice of vertices, see Eq.\ (\ref{axes}).}}
\end{figure}

Artificial spin ice on a square lattice \cite{Skjaervo19} has attracted a lot of attention as a system exhibiting
frustration \cite{Nisoli13,Nguyen17}, a doubly-degenerate ground state \cite{Morgan11}, and monopole-like topological 
excitations \cite{Silva12,Ribeiro+17,Nisoli18} out of a degenerate ground state \cite{Moller06,Moller09,Mol09}.  
These properties result from the geometry-induced demagnetization anisotropy of magnetic islands fabricated on a
nonmagnetic substrate \cite{Wang06}. 
Protocols using applied magnetic fields have been developed to nudge the system towards a ground 
state \cite{Ke08,Nisoli10,Porro+13,Farhan13,Zhang+19}, which is difficult due to the energy barriers associated 
with frustration.
\change{In a ground state ({\sc gs}) of artificial square ice, the island dipoles satisfy the two-in, two-out ice rule at 
each vertex to minimize the dipolar energy, also refered to as Type I, with pairs of opposing dipoles both inward 
or both outward across the centers of the vertices, see Fig.\ 1 of Ref.\ \onlinecite{Kapaklis+12}.}
If the system can be pushed into a ground state, it should have a particular small-amplitude
spin wave spectrum, which has been investigated in varying approximations \cite{Gliga+13,Jung+16,Iacocca+16}.
The spectrum is expected to help identify and characterize the ground state \cite{Arroo+19,Lasnier+20}.

A remanent state ({\sc rs}) of spin ice, however, may be more straightforward to obtain, as it 
requires application of an applied field that is slowly reduced to zero. 
Fig.\ \ref{square-rs} shows a segment of square ice left in a remanent state, after applying
a field along \change{a nearest-neighbor ({\sc nn}) primary axis of the island lattice, labeled} 
$x_{\rm I}$, which was then turned off. 
The state is metastable, being a local energy minimum but well above the ground state, 
\change{and although it also satisfies the ice rule, the vertices are of higher energy and refered to
as Type II \cite{Kapaklis+12} vertices.}

Remanent states should possess distinct small-amplitude oscillations or spinwave spectrum, that 
signals the presence of that state \cite{Arroo+19,Arora+Das21}.
The goal of the present work is to estimate the stability properties of a remanent state,
by analysis of the linearized spin wave modes about a remanent state, by assuming Heisenberg-like
dipole dynamics \cite{Heisenberg28,Wysin+13} as opposed to Ising spins \cite{Ising25,Ostman18}.
\change{There is one dipole per island, of fixed length but varying direction. This assumption ignores 
internal magnetization dynamics within the islands.  For isolated thin islands with large in-plane 
aspect ratios, simulations show that there is very little spatial variation in the internal magnetization, even
under a reversal process \cite{Wysin+12}. This approximation will be valid if the dipolar interaction fields 
are nearly uniform within an island affected by those fields.}

\change{The Heisenberg-like island dipoles in the {\sc rs} of Fig.\ \ref{square-rs} are tilted slightly
from the islands' long axes. This is because dipolar interactions cause the dipoles on the two sublattices 
to tilt towards each other as they compete with the shape anisotropy of the islands. The effects of this 
tilting are taken into account here.}

Long-range dipole interactions have been shown to be highly relevant \cite{Brunn+21}.
For clarity, we start from a {\sc nn} model and extend it to include all
dipolar interactions to unlimited range.
The sum over infinite-range dipole interactions is motivated by a calculation of the mode spectrum for a 
one-dimensional chain of magnetic islands \cite{Wysin21}.
The mode spectrum helps to determine the stability properties and discriminates remanent states from other configurations.

\subsection{Heisenberg-like dipole model}
In this model \cite{Wysin+13} the magnetic islands have single-domain dipole moments of fixed magnitude $\mu$ whose time-dependent 
directions are along Heisenberg-like unit spin vectors $\bm{\hat\mu}_i(t)$. 
The elongation of the islands produces uniaxial anisotropy \cite{Wysin+12} along axis $\bm{\hat{u}}_i$ of strength $K_1$, and 
their limited height produces planar anisotropy of strength $K_3$ with axis $\bm{\hat{z}}$.  
Both anisotropies are due to demagnetization or geometric effects.
\change{The anisotropy energies are very close to parabolic in the components of the dipole \cite{Wysin+12},
even though small deviations from uniform magnetization can appear at the edges.}
The islands are elongated either along the $x$ or $y$ directions of a square lattice of vertices, and they are 
symmetrically located between vertices at locations $(v_x, v_y)a_{\rm v}$, where the spacing is 
$a_{\rm v}$ and $v_x,v_y$ are integer locations.
The Hamiltonian for $N$ islands can be written
\bn
\label{Ham}
{H} &=& -\frac{\mu_0}{4\pi} \frac{\mu^2}{a_{\rm I}^3}
\sum_{i>j}^N \frac{ \left[ 3(\bm{\hat\mu}_i\cdot {\bf \hat{r}}_{ij})(\bm{\hat\mu}_j\cdot{\bf \hat{r}}_{ij})
                                -\bm{\hat\mu}_i\cdot \bm{\hat\mu}_j \right]}
{\left( {r}_{ij} / a_{\rm I}\right)^3} \nonumber \\
&+&  \sum_{i} \left\{ K_1[1-(\bm{\hat\mu}_{i}\cdot{\bf \hat{u}}_i)^2]
+ K_3 (\bm{\hat\mu}_{i}\cdot {\bf \hat{z}})^2 \right\}
\en
where $\mu_0$ is the magnetic permeability of space, $a_{\rm I}$ is the {\sc nn}-spacing of the islands
and $r_{ij}$ and ${\bf\hat{r}}_{ij}$ are \change{center-to-center} distance and direction vectors between pairs of islands \cite{Wysin+15}.
The {\sc nn} island separation is $a_{\rm I}=\frac{1}{\sqrt{2}}a_{\rm v}$, which determines the {\sc nn}
principal displacements (i.e., basis vectors of the island lattice), 
\change{
\be
\label{axes}
{\bf x}_{\rm I}=\tfrac{1}{\sqrt{2}}a_{\rm I} (\hat{\bf x}+\hat{\bf y}), \quad
{\bf y}_{\rm I}=\tfrac{1}{\sqrt{2}}a_{\rm I} (-\hat{\bf x}+\hat{\bf y}),
\ee
}
rotated 45$^{\circ}$ from the $xy$ coordinate system of the vertices, see Fig.\ \ref{square-rs}.
\change{When indicating directions in this work, the island {\sc nn} principal directions along $\hat{\bf x}_{\rm I}$ 
and $\hat{\bf y}_{\rm I}$ are used.  For example, the net magnetization of the state in Fig.\ \ref{square-rs}
is along the [10] direction of the island lattice (equivalent to the [11] direction of the vertex lattice).}

A convenient energy unit is the {\sc nn} dipolar coupling, denoted with script ${\cal D}$,
\be
\label{Dd}
{\cal D} \equiv \frac{\mu_0}{4\pi}\frac{\mu^2}{a_{\rm I}^3},
\ee
and farther neighbors have dipole interactions reduced by the \change{center-to-center} distance cubed.

\section{Remanent states in the {\sc nn}-model}
\label{nn-model}

\begin{figure}
\includegraphics[width=\figwidth,angle=0]{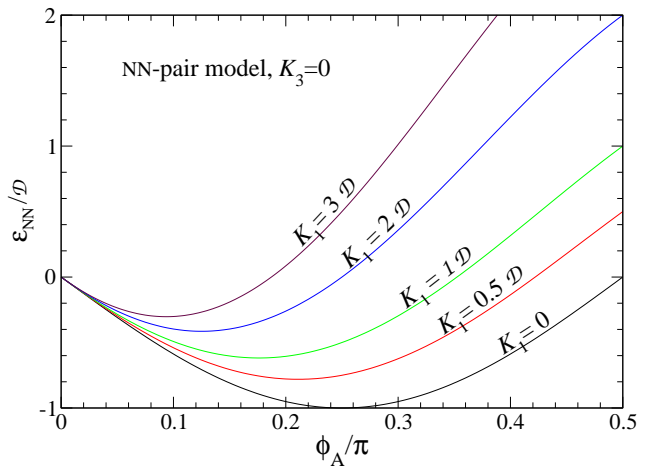}
\caption{\label{Uphi}
The {\sc rs} energy $\varepsilon_{\rm NN}(\phi_A)$ given in Eq.\ (\ref{Uf}) for various values of
in-plane anisotropy, $K_1/{\cal D}$.  In all cases there is a local minimum, which suggests
real frequencies of oscillation about that minimum, however, the {\sc nn}-model requires 
$K_1 > 2.947 {\cal D}$ for stability, which limits the stable canting angle $\phi_A$.}
\end{figure}

First we consider only {\sc nn} dipole interactions, and later include long-range interactions that are 
known to be important \cite{Rougemaille11, Shevchenko17} in the dynamics of spin ice.
The system is assumed to be uniform by sublattice.
The islands aligned along ${\bf\hat{x}}$ make up the A sublattice, with spins $\bm{\hat{\mu}}_i$={\bf A},
 and the islands aligned along ${\bf\hat{y}}$ comprise the B sublattice, with spins $\bm{\hat{\mu}}_i$={\bf B}.
A central A-site interacts with four {\sc nn} B-sites, and {\it vice-versa}. The dipole interactions depend on the direction to 
the neighbors.

Taking the interaction of an A-site with its neighbors, averaged with the interaction of a B-site with its neighbors,
leads to an effective two-sublattice Hamiltonian, which is the energy per pair of A and B sites,
\begin{align}
H_{\rm AB} &= -2{\cal D} \left[ 
3({\bf A}\cdot{\bf \hat{x}_I})({\bf B}\cdot{\bf \hat{x}_I}) 
+3({\bf A}\cdot{\bf \hat{y}_I})({\bf B}\cdot{\bf \hat{y}_I})  \right. \nonumber \\
  -2& \left. {\bf A}\cdot{\bf B}\right] 
 +K_1\left(2-A_x^2-B_y^2\right)+K_3\left(A_z^2+B_z^2\right).
\end{align}
Inserting the {\sc nn} unit vectors, this is
\bn
H_{\rm AB} &=& -2{\cal D} \left(A_x B_x +A_y B_y-2A_z B_z\right) \nonumber \\
&+&K_1\left(2-A_x^2-B_y^2\right)+K_3 \left(A_z^2+B_z^2\right).
\en
A local minimum of this Hamiltonian is a remanent state.  It should be minimized under the constraint of
fixed spin length for ${\bf A}$ and ${\bf B}$. Lagrange's method of undetermined multipliers
quickly shows that $A_z=B_z=0$ is required; the dipoles remain within the $xy$-plane.    
Thus their equilibrium directions are described by in-plane angles $\phi_A$ and $\phi_B$, 
taken as counterclockwise deviations away from the ${\bf \hat{x}}$ and ${\bf \hat{y}}$ directions,
such that,
\be
{\bf A} = (\cos\phi_A, \sin\phi_A, 0), \ \ 
{\bf B} = (-\sin\phi_B, \cos\phi_B, 0). 
\ee
In these coordinates the two-sublattice Hamiltonian is
\be 
H_{\rm AB}=-2{\cal D} \sin(\phi_A-\phi_B) +K_1(2-\cos^2\!\phi_A-\cos^2\!\phi_B).
\ee 
This is minimized when $\sin 2\phi_A = -\sin 2 \phi_B$, \change{which implies $\phi_B=-\phi_A$,
together} with the additional requirement, 
\change{
\be
\frac{\partial H_{\rm AB}}{\partial \phi_A} = -2{\cal D} \cos(\phi_A-\phi_B)+K_1 \sin 2\phi_A =0.
\ee
}
The fourfold symmetry of the system implies four {\sc rs} solutions.
We consider the primary one as that where the system magnetization points at 45$^{\circ}$ 
between the $x$ and $y$ axes (along ${\bf x}_{\rm I}$), with energy-minimizing angles satisfying
\be
\label{RS0}
\tan 2\phi_A^{0} = -\tan 2\phi_B^{0} = \frac{2{\cal D}}{K_1} .
\ee
The $0$ superscript indicates the equilibrium {\sc rs} values.
With $\phi_A^{0}$ positive, and $\phi_B^{0}$ negative, \change{as in Fig.\ \ref{square-rs},} the sublattices tilt 
inward towards the 45$^{\circ}$ diagonal direction, attempting to minimize dipolar energy which competes with an 
increasing uniaxial anisotropy energy. 
The inward canting of the sublattices is small, unless $K_1$ is small.
However, when $K_1$ is too small the state will destabilize, as is shown later.
Thus there is a limited amount of spin canting.

\begin{figure}
\includegraphics[width=\figwidth,angle=0]{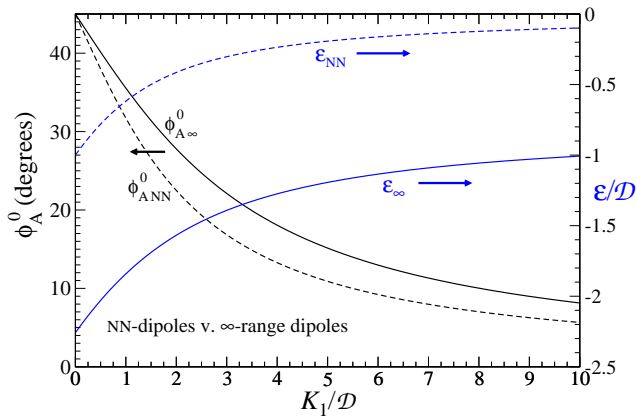}
\caption{\label{eps+phiA}
The {\sc rs} equilibrium tilting angles $\phi_A^0$ (left axis) given in Eqs.\ \ref{RS0}, and the equilibrium energy
per site $\varepsilon$ (right axis) in Eqs.\ \ref{Uf}, as functions of the islands' uniaxial anisotropy
$K_1/{\cal D}$, comparing the nearest-neighbor couplings model with the infinite-range couplings model \change{in
Sec.\ \ref{lrdeq}}.}
\end{figure}
With $\phi_B=-\phi_A$, the energy per site is
\be
\label{Uf}
\varepsilon_{\rm NN}(\phi_A) = \tfrac{1}{2}H_{\rm AB} = -{\cal D}\sin 2\phi_A+K_1 \sin^2\!\phi_A.
\ee
This is minimized when $\phi_A=\phi_A^{0}$, as seen in Fig.\ \ref{Uphi} for various anisotropy strengths relative 
to dipole strength.
Stronger uniaxial anisotropy in the islands reduces the spin canting, see Fig.\ \ref{eps+phiA}, which
shows the canting angle $\phi_A^0$ and energy per site $\varepsilon_{\rm NN}$ for the {\sc nn}-model and for the 
model that includes infinite-range dipole interactions, below.
For comparison, the per-site energy in a ground state of square ice is $\varepsilon_{\rm gs} = -3{\cal D}$
when only {\sc nn}-interactions are included.
Later we show that the {\sc nn}-model requires $K_1 > 2.947\, {\cal D}$ for {\sc rs} stability, hence the {\sc nn}
canting angle is limited by $\phi_{A}^{0} < 17^{\circ}$.
Even so, this angular deviation must be taken into account to obtain 
correct normal mode oscillation frequencies around a remanent state.
%

\section{Linearized oscillation dynamics in the {\sc nn}-model}
Still staying with the {\sc nn}-model, the small amplitude oscillations can be analyzed.
The island dipoles are now allowed to deviate slightly from their equilibrium {\sc rs} directions 
in Eq.\ (\ref{RS0}), including out-of-plane deviations $\mu_{\bf n}^z=\mu \sin\theta_{\bf n}$.
An island is located at 
\be
{\bf n}=n_{x_{\rm I}} {\bf x}_{\rm I} + n_{y_{\rm I}}  {\bf y}_{\rm I},
\ee
in terms of the {\sc nn} principal displacements ${\bf x}_{\rm I}, {\bf y}_{\rm I}$, 
letting $n_{x_{\rm I}}=1,2,3,...N_1$ and $n_{y_{\rm I}}=1,2,3,...N_2$.
The total energy is a sum over islands {\bf n}, and counting 
the dipole interactions with a {\sc nn} bond to ${\bf n+x_I}$
and a {\sc nn} bond to ${\bf n+y_I}$.
The dipole interaction energies in these different directions are not equivalent.

Based on (\ref{Ham}), the dipole-dipole interaction energy of an island at {\bf n} on the A-sublattice with
the neighbor at ${\bf n+x_I}$ on the B-sublattice is
\bn
u_{\rm dd_{\parallel}} &=& -{\cal D} \left[ \tfrac{3}{2}(A_{\bf n}^x B_{\bf n+x_I}^y +A_{\bf n}^y B_{\bf n+x_I}^x) \right. \\
&+&\left. \tfrac{1}{2} (A_{\bf n}^x B_{\bf n+x_I}^x +A_{\bf n}^y B_{\bf n+x_I}^y) -A_{\bf n}^z B_{\bf n+x_I}^z \right].
\nonumber
\en
The parallel symbol $(\parallel)$ indicates that the spins point close to the bond direction.
For the interaction of an A-site with its neighbor at ${\bf n+y_I}$ on the B-sublattice, the first
term has the opposite sign,
\bn
u_{\rm dd_{\perp}} &=& -{\cal D} \left[-\tfrac{3}{2}(A_{\bf n}^x B_{\bf n+y_I}^y +A_{\bf n}^y B_{\bf n+y_I}^x) \right. \\
&+&\left. \tfrac{1}{2} (A_{\bf n}^x B_{\bf n+y_I}^x +A_{\bf n}^y B_{\bf n+y_I}^y) -A_{\bf n}^z B_{\bf n+y_I}^z \right].
\nonumber
\en
The perpendicular symbol $(\perp)$ indicates that the spins point almost perpendicular to the bond direction.
These expressions also apply to the interaction of a B-island with its principal direction neighbors
on the A-sublattice (interchanging A and B).

To analyze time-dependent fluctuations, the in-plane angles are set to $\phi_A=\phi_A^0+\phi_{\bf n}(t)$ on A-islands and 
$\phi_B=\phi_B^0+\phi_{\bf n}(t)$ on B-islands, where $\phi_{\bf n}(t)$ are the deviations from the equilibrium {\sc rs}.
There are nonzero out-of-plane deviations $\theta_{\bf n}(t)$, such that the spins' $(x,y,z)$  components are written
\bn
\label{AB-devs}
{\bf A}_{\bf n} &=& (\cn \cos(\phi_A^0+\phi_{\bf n}), \cn \sin(\phi_A^0+\phi_{\bf n}), \sn),  \\
{\bf B}_{\bf n} &=& (-\cn \sin(-\phi_A^0+\phi_{\bf n}), \cn \cos(-\phi_A^0+\phi_{\bf n}), \sn), \nonumber
\en
where $\cn=\cos\theta_{\bf n}, \ \sn=\sin\theta_{\bf n}$. 
\begin{widetext}
\noindent
Then, the dipolar energies can be expanded to quadratic order in $\phi_{\bf n}\ll 1$ and $\theta_{\bf n}\ll 1$.
For example, 
\begin{align}
\frac{u_{\rm dd_{\parallel}}}{\cal D} & \approx
-\tfrac{3}{2}\left[1-\phi_{\bf n}\phi_{\bf n+x_I}-\tfrac{1}{2}\left(\phi_{\bf n}^2+\phi_{\bf n+x_I}^2
+\theta_{\bf n}^2+\theta_{\bf n+x_I}^2\right)\right] +\theta_{\bf n}\theta_{\bf n+x_I}  \nonumber \\
& -\tfrac{1}{2}\left\{ \sin(2\phi_A^0)\left[1+\phi_{\bf n}\phi_{\bf n+x_I}-\tfrac{1}{2}\left(\phi_{\bf n}^2 +\phi_{\bf n+x_I}^2
+\theta_{\bf n}^2+\theta_{\bf n+x_I}^2\right)\right] +\cos(2\phi_A^0)(\phi_{\bf n}-\phi_{\bf n+x_I})\right\}.
\end{align}
A similar expression gives $u_{\rm dd_{\perp}}$, with the $-\frac{3}{2}$ changed to $+\frac{3}{2}$.
Combining $u_{\rm dd_{\parallel}}$ with $u_{\rm dd_{\perp}}$, and summing over ${\bf n}$ produces the net
{\sc nn} dipolar energy, ordered by zeroth, linear, and quadratic terms,
\begin{align}
U_{\rm dd} & \approx -N{\cal D}\sin(2\phi_A^0) 
-{\cal D}\cos(2\phi_A^0) \sum_{\bf n} \left[\phi_{\bf n}-\tfrac{1}{2}(\phi_{\bf n+x_I}+\phi_{\bf n+y_I})\right] \nonumber \\
& +{\cal D}\sum_{\bf n} \left\{ \tfrac{3}{2}\phi_{\bf n}(\phi_{\bf n+x_I}-\phi_{\bf n+y_I})
+\theta_{\bf n}(\theta_{\bf n+x_I}+\theta_{\bf n+y_I}) 
+\sin(2\phi_A^0)\left[\phi_{\bf n}^2+\theta_{\bf n}^2-\tfrac{1}{2}\phi_{\bf n}(\phi_{\bf n+x_I}+\phi_{\bf n+y_I})\right] \right\}.
\end{align}
In the same way, the anisotropy energy after expansion is
\begin{align}
\label{UK}
U_K & \approx 
N K_1 \sin^2\!\phi_A^0 
+ \tfrac{K_1}{2}\sin(2\phi_A^0) \sum_{\bf n} \left[\phi_{\bf n}-\tfrac{1}{2}(\phi_{\bf n+x}+\phi_{\bf n+y})\right]
+\sum_{\bf n} \left[ K_1 \cos(2\phi_A^0)\phi_{\bf n}^2+\left(K_1 \cos^2\phi_A^0+K_3\right)\theta_{\bf n}^2 \right].
\end{align}
\end{widetext}
The total system energy is the sum,
\be
H = U_{\rm dd}+U_K = H^{0}+H^{(1)}+H^{(2)},
\ee
where the zeroth order term $H^{(0)}$ is the {\sc rs} energy: 
\be
H^{(0)} = N\left(-{\cal D}\sin 2\phi_A^0 +K_1\sin^2\!\phi_A^0\right).
\ee
The term $H^{(1)}$ linear in deviations is zero,
and the quadratic terms are separated into in-plane parts and out-of-plane parts, $H^{(2)}=H_{\phi}+H_{\theta}$,
defined by 
\begin{align}
\label{Hfnn}
H_{\phi}  = \sum_{\bf n} & \left\{ \left[{\cal D}\sin 2\phi_A^0 +K_1\cos 2\phi_A^0 \right]\phi_{\bf n}^2
\right. \nonumber \\
& -\tfrac{1}{2}{\cal D}\sin(2\phi_A^0)~ \phi_{\bf n}(\phi_{\bf n+x_I}+\phi_{\bf n+y_I})
\nonumber \\
& \left. +\tfrac{3}{2}{\cal D}\phi_{\bf n}(\phi_{\bf n+x_I}-\phi_{\bf n+y_I}) \right\},
\end{align}
\begin{align}
\label{Htnn}
H_{\theta}  = \sum_{\bf n} & \left\{ \left[{\cal D}\sin 2\phi_A^0 +K_1 \cos^2\!\phi_A^0+K_3\right]\theta_{\bf n}^2
\right. \nonumber \\
& \left. +{\cal D}\theta_{\bf n}(\theta_{\bf n+x_I}+\theta_{\bf n+y_I}) \right\} .
\end{align}
%

\subsection{Spin deviation energy in matrix form}
Having expressed the small fluctuations by quadratic Hamiltonians, now it is possible to extract
the modes of oscillation.
To that end, the sub-Hamiltonians $H_{\phi}$ and $H_{\theta}$ can be written in a matrix notation, 
from which the spin dynamics is easier to follow.
Generally, the spin deviations form state vectors (written as row vectors),
\begin{align}
\psi_{\phi}^{\dagger} &=(\phi_1,\phi_2,\phi_3,...\phi_N), \nonumber \\
\psi_{\theta}^{\dagger} &=(\theta_1,\theta_2,\theta_3,...\theta_N), 
\end{align}
where the subscripts label the islands.
There are only sparse couplings (i.e., nearest neighbors) among the elements of each vector. 
The out-of-plane Hamiltonian can be written in terms of an $N\times N$ matrix $\bm{M}_{\theta}$ as
\be
H_{\theta}=\tfrac{1}{2} \psi_{\theta}^{\dagger} \bm{M}_{\theta} \psi_{\theta},
\ee
where the matrix elements are either diagonal ones ($M_{\theta,{\bf n,n}}$) or {\sc nn} ones 
($M_{\theta,{\bf n,n \pm x_I}}$ and $M_{\theta,{\bf n,n \pm y_I}}$).
From (\ref{Htnn}), the nonzero elements are
\begin{align} 
\label{Mt}
M_{\theta,{\bf n,n}}  & =  M_{\theta,1} \equiv 2\left({\cal D}\sin 2\phi_A^0 +K_1 \cos^2\!\phi_A^0+K_3\right), \nonumber \\
M_{\theta,{\bf n, n \pm x_I}} & = M_{\theta,{\bf n, n \pm y_I}}  \equiv M_{\theta,2} = {\cal D}.
\end{align}
$M_{\theta}$ is symmetric: $M_{\theta,{\bf n,m}}=M_{\theta,{\bf m,n}}$. 
The in-plane Hamiltonian can be written in the same form, 
\be
H_{\phi}=\tfrac{1}{2}\psi_{\phi}^{\dagger} \bm{M}_{\phi}\psi_{\phi},
\ee
where (\ref{Hfnn}) gives the nonzero matrix elements,
\begin{align} 
\label{Mf}
M&_{\phi,{\bf n,n}} = M_{\phi,1}  \equiv 2\left({\cal D}\sin 2\phi_A^0+K_1 \cos 2\phi_A^0\right), \nonumber \\
M&_{\phi,2} \equiv -\tfrac{1}{2}{\cal D}\sin 2\phi_A^0, \quad  M_{\phi,3}  \equiv \tfrac{3}{2}{\cal D},  \nonumber \\
M&_{\phi,{\bf n, n \pm x_I}} = M_{\phi,2}+M_{\phi,3}, \nonumber \\
M&_{\phi,{\bf n, n \pm y_I}} = M_{\phi,2}-M_{\phi,3}.
\end{align} 
$\bm{M}_{\phi}$ is also symmetric,  but the couplings in the ${\bf y_I}$ direction are different than those in 
the ${\bf x_I}$ direction.

\subsection{Spin dynamics from $H_{\phi}$ and $H_{\theta}$}
Assuming a gyromagnetic ratio $\gamma_{\rm e}$, the undamped dynamics of an island's  magnetic moment is
given by a torque equation \cite{Jiles91,Wysin15},
\be
\label{torq}
\frac{d \bm{\mu}_{\bf n}}{dt} = \gamma_{\rm e} \bm{\mu}_{\bf n} \times {\bf B}_{\bf n}.
\ee
The magnetic field that acts on an island is derived from the total Hamiltonian, $H=H_{\phi}+H_{\theta}$,
according to 
\be
{\bf B}_{\bf n} = -\frac{\partial H}{\partial \bm{\mu}_{\bf n}}.
\ee
Each island's unit spin $\bm{\hat{\mu}}_{\bf n}$ is nearly aligned to the equilibrium magnetic field 
${\bf B}^0_{\bf n}$ acting on it, except for the small deviations caused by oscillations.
The fluctuations in ${\bf B}_{\bf n}$ contribute to torque.
Let $\bm{\hat{\mu}}_{\bf n}^0$ be the equilibrium spin, and then let
$\bm{\hat{t}}_{\bf n}=\bm{\hat{z}}\times \bm{\hat{\mu}}_{\bf n}^0$ be a unit vector transverse to  $\bm{\hat{\mu}}_{\bf n}^0$
in the $xy$-plane.
The spin has transverse deviation $\phi_{\bf n} \bm{\hat{t}}_{\bf n}$ and $z$-deviation $\theta_{\bf n} \bm{\hat{z}}$,
such that
\be
\bm{\hat{\mu}}_{\bf n} \approx \bm{\hat{\mu}}_{\bf n}^0+\phi_{\bf n}\bm{\hat{t}}_{\bf n} + \theta_{\bf n} \bm{\hat{z}}.
\ee
The magnetic field can also be expressed as the equilibrium value plus transverse and $z$-deviations,
\be
{\bf B}_{\bf n} = {\bf B}_{\bf n}^0-\frac{1}{\mu}\frac{\partial H}{\partial\phi_{\bf n}} \bm{\hat{t}}_{\bf n}
-\frac{1}{\mu}\frac{\partial H}{\partial\theta_{\bf n}} \bm{\hat{z}}.
\ee
Inserting into the torque equation (\ref{torq}), and keeping only leading terms, gives 
\be
\dot{\phi}_{\bf n}\bm{\hat{t}}_{\bf n} + \dot{\theta}_{\bf n} \bm{\hat{z}} = \gamma_e  \bm{\hat{\mu}}_{\bf n}^0
\times \left( {\bf B}_{\bf n}^0-\frac{1}{\mu}\frac{\partial H}{\partial\phi_{\bf n}} \bm{\hat{t}}_{\bf n}
-\frac{1}{\mu}\frac{\partial H}{\partial\theta_{\bf n}} \bm{\hat{z}}\right).
\ee
Separating into transverse and $z$-components gives the linearized Hamiltonian equations of motion,
\be
\label{dots}
\dot{\phi}_{\bf n}  = \frac{\gamma_{\rm e}}{\mu} \frac{\partial H}{\partial \theta_{\bf n}}, \quad
\dot{\theta}_{\bf n} = -\frac{\gamma_{\rm e}}{\mu} \frac{\partial H}{\partial \phi_{\bf n}}.
\ee
These result more directly by realizing that \change{out-of-plane component 
$\bm{\hat{\mu}}_{\bf n}^z = \theta_{\bf n}$} is the momentum conjugate to $\phi_{\bf n}$.

With $H=H_{\phi}+H_{\theta}$ in separated form, the set of derivatives can be expressed
via matrix notation,
\be
\left(\frac{\partial H_{\phi}}{\partial \phi_{\bf n}}\right) = \bm{M}_{\phi} \psi_{\phi}, \quad
\left(\frac{\partial H_{\theta}}{\partial \theta_{\bf n}}\right) = \bm{M}_{\theta} \psi_{\theta}.
\ee
The dynamic equations become a matrix problem with $2N$ degrees of freedom,
\be
\label{matrix}
\dot{\psi}_\phi = \frac{\gamma_e}{\mu}  \bm{M}_{\theta} \psi_{\theta}, \quad
\dot{\psi}_\theta = -\frac{\gamma_e}{\mu}  \bm{M}_{\phi} \psi_{\phi}. 
\ee

There are various ways to solve (\ref{matrix}) for the dynamic eigenmodes. One
way that works only in the {\sc nn}-model is to find the eigenvalues and eigenvectors
of $\bm{M}_{\theta}$ and $\bm{M}_{\phi}$, that satisfy 
\be
\bm{M}_{\theta}\psi_{\theta}=\lambda_{\theta} \psi_{\theta}, \quad
\bm{M}_{\phi}\psi_{\phi}=\lambda_{\phi} \psi_{\phi}.
\ee

\subsubsection{Eigenvalues of $\bm{M}_{\theta}$}
%
\begin{figure}
\includegraphics[width=\figwidth,angle=0]{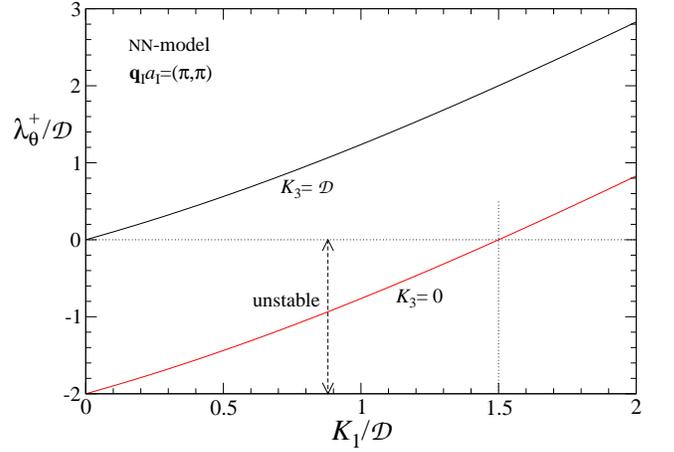}
\caption{\label{lam-theta} Energy eigenvalue $\lambda_{\theta}^{+}$ vs.\  in-plane anisotropy, 
for the most unstable wave vector ${\bf q}_{\rm I}a_{\rm I}=(\pi,\pi)$ \change{in island coordinates},
showing that stability requires $K_1>1.5{\cal D}$ when $K_3=0$, or any $K_1>0$ when $K_3>{\cal D}$.
The same results hold for $\lambda_{\theta}^{-}(0,0)$.}
\end{figure} 

Consider $\bm{M}_{\theta}$. The elements $\theta_{\bf n}$ of an eigenvector are 
identified by their position, ${\bf n}=n_{x_{\rm I}} {\bf x}_{\rm I}+n_{y_{\rm I}}{\bf y}_{\rm I}$, 
in the {\sc nn} island coordinates.
One row of the eigenvalue problem for $\bm{M}_{\theta}$ is
\begin{align}
\label{row}
M_{\theta,1}\theta_{\bf n} &+ M_{\theta,2}(\theta_{\bf n+x_{\rm I}}+\theta_{\bf n-x_{\rm I}})
\nonumber \\
& +M_{\theta,2}(\theta_{\bf n+y_{\rm I}} +\theta_{\bf n-y_{\rm I}}) = \lambda_{\theta}~ \theta_{\bf n},
\end{align}
where the on-site ($M_{\theta,1}$) and {\sc nn} ($M_{\theta,2}$) matrix elements were defined in (\ref{Mt}).
The out-of-plane fluctuations for a dipole are 
$\mu_{\bf n}^z=\mu\sin\theta_{\bf n}\approx \mu \theta_{\bf n}$.

With periodic boundary conditions, wave solutions result.
The elements of $\psi_{\theta}$ are either on the A or B sublattice. Thus two
amplitudes $a_{\theta}, b_{\theta}$ are included, {\it i.e.},  
\bn
\theta_{\bf n}^{\rm A} &=& a_{\theta} e^{i {\bf q}\cdot {\bf n}},  \ \text{A-sites},
\nonumber \\
\theta_{\bf n}^{\rm B} &=& b_{\theta} e^{i {\bf q}\cdot {\bf n}}, \ \text{B-sites}.
\en
The wave vectors ${\bf q}=(q_{x_{\rm I}}, q_{y_{\rm I}})$ are quantized in the usual way, with components
in island coordinates,
\bn
\label{qvec}
q_{x_{\rm I}} &=& \frac{2\pi k_x}{N_1 a_{\rm I}},\ k_x=0,1,2,...(N_1-1), \nonumber \\
q_{y_{\rm I}} &=& \frac{2\pi k_y}{N_2 a_{\rm I}},\ k_y=0,1,2,...(N_2-1).
\en
For the wave solution, the sums over {\sc nn}'s are
\begin{align}
\theta_{\bf n+x_{\rm I}}+\theta_{\bf n-x_{\rm I}}&=2\theta_{\bf n}\cos q_{\bf x_I} a_{\rm I}, \nonumber \\
\theta_{\bf n+y_{\rm I}}+\theta_{\bf n-y_{\rm I}}&=2\theta_{\bf n}\cos q_{\bf y_I} a_{\rm I}. 
\end{align}
The total {\bf q}-dependent phase factor is
\be
\gamma_{\bf q}^{+} = 2(\cos q_{x_{\rm I}} a_{\rm I}+\cos q_{y_{\rm I}} a_{\rm I})
\ee
The system of equations (\ref{row}) is split into two sets, depending on whether ${\bf n}$ resides 
on the A or B sublattice, producing a $2\times 2$ reduced system:
\be
\left[ \begin{array}{cc}
M_{\theta,1} & M_{\theta,2} \gamma_{\bf q}^{+}  \\ M_{\theta,2} \gamma_{\bf q}^{+}  & M_{\theta, 1} \end{array} \right]
 \left[ \begin{array}{c} a_{\theta} \\ b_{\theta} \end{array} \right]
 = \lambda_{\theta} \left[ \begin{array}{c} a_{\theta} \\ b_{\theta} \end{array} \right].
\ee
This matrix has symmetric and antisymmetric eigenvectors, 
$(a_{\theta},b_{\theta})=(\psi^{\pm})^{\dagger} \equiv \frac{1}{\sqrt{2}}(1,\pm 1)$, with eigenvalues
\be
\lambda_{\theta}^{\pm} = M_{\theta,1}\pm M_{\theta,2}\gamma_{\bf q}^{+}.
\ee
In terms of the original constants in $H$ and with full ${\bf q}$-dependence this is
\begin{align}
 \lambda_{\theta}^{\pm}({\bf q}) = &  2\left[{\cal D}\sin 2\phi_A^0+K_1 \cos^2\!\phi_A^0+K_3 \right.
\nonumber \\
 & \left. \pm {\cal D}\left(\cos q_{x_{\rm I}} a_{\rm I}+\cos q_{y_{\rm I}} a_{\rm I}\right)\right].  
\end{align}
A negative eigenvalue implies that the {\sc rs} can lower its energy by excitation with the associated
wave vector, which indicates instability.  Specifically of interest, $\lambda_{\theta}^{+}(\pi,\pi)$ is 
plotted in Fig.\ \ref{lam-theta}, which shows {\sc rs} instability for $K_1>1.5{\cal D}$ when $K_3=0$, and 
stability for all $K_1>0$ when $K_3>{\cal D}$.  \change{Note that wave vectors are denoted in terms of their
components and directions in the island coordinates.}

\subsubsection{Eigenvalues of $\bm{M}_{\phi}$}
Next, consider $\bm{M}_{\phi}$, which has the matrix elements in (\ref{Mf}).  
Fluctuations in $\phi_{\bf n}$ are dipole components {\em transverse} to the equilibrium 
directions, within the $xy$-plane.  
One row of the eigenvalue problem is 
\begin{align}
\label{Hdyn}
M_{\phi,1}\phi_{\bf n} & +(M_{\phi,2}+M_{\phi,3})(\phi_{\bf n+x_I}+\phi_{\bf n-x_I})
\nonumber \\
& +(M_{\phi,2}-M_{\phi,3})(\phi_{\bf n+y_I}+\phi_{\bf n-y_I}) =  \lambda_{\phi} \phi_{\bf n}.
\end{align}
Note that couplings along the ${\bf y_I}$ direction are different than along ${\bf x_I}$.
Again a wave solution is present, with amplitudes $a_{\phi}, b_{\phi}$ on the sublattices, 
{\it i.e.},
\bn
\phi_{\bf n}^{\rm A} &=& a_{\phi} e^{i {\bf q}\cdot {\bf n}},  \ \text{A-sites},
\nonumber \\
\phi_{\bf n}^{\rm B} &=& b_{\phi} e^{i {\bf q}\cdot {\bf n}}, \ \text{B-sites}.
\en
\begin{figure}
\includegraphics[width=\figwidth,angle=0]{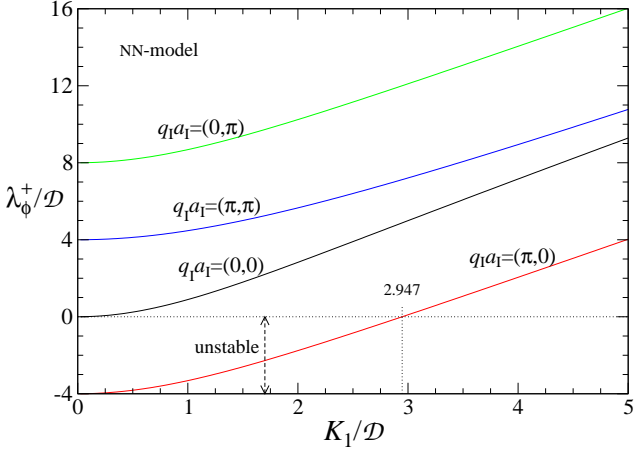}
\caption{\label{lam-phi} The {\sc nn}-model in-plane energy eigenvalue $\lambda_{\phi}^{+}$ vs. in-plane anisotropy, for
various wave vectors \change{in island coordinates,} Eq.\ \ref{lamphi}, showing that stability requires $K_1>K_{1,\rm min}\approx 2.947\, {\cal D}$.}
\end{figure}
Defining another phase factor,
\be
\gamma_{\bf q}^{-} = 2(\cos q_{x_{\rm I}} a_{\rm I}-\cos q_{y_{\rm I}} a_{\rm I}),
\ee
the reduced $2\times 2$ system is
\begin{align}
&\left[ \begin{array}{cc} n_{aa} & n_{ab}  \\ n_{ba} & n_{bb} \end{array} \right]
 \left[ \begin{array}{c} a_{\phi} \\ b_{\phi} \end{array} \right] 
 = \lambda_{\phi} \left[ \begin{array}{c} a_{\phi} \\ b_{\phi} \end{array} \right].  \nonumber \\
&n_{aa}=n_{bb} \equiv M_{\phi,1}, \nonumber \\ 
&n_{ab}=n_{ba} \equiv M_{\phi,2} \gamma_{\bf q}^{+}+M_{\phi,3}\gamma_{\bf q}^{-}.
\end{align}
The eigenvectors are again symmetric and antisymmetric, 
$(a_{\phi},b_{\phi})=(\psi^{\pm})^{\dagger}=\frac{1}{\sqrt{2}}(1,\pm 1)$ with eigenvalues
\be
\lambda_{\phi}^{\pm} = M_{\phi,1}\pm \left(M_{\phi,2}\gamma_{\bf q}^{+}+M_{\phi,3}\gamma_{\bf q}^{-}\right).
\ee
The full dependence on ${\bf q}$ and parameters in $H$ is
\begin{align}
\label{lamphi}
\lambda_{\phi}^{\pm}({\bf q}) & = 2K_1 \cos 2\phi_A^0 
 \pm 3{\cal D}(\cos q_{x_{\rm I}} a_{\rm I}- \cos q_{y_{\rm I}} a_{\rm I}) \nonumber \\
& +{\cal D}\sin 2\phi_A^0 [2 \mp(\cos q_{x_{\rm I}} a_{\rm I}+\cos q_{y_{\rm I}} a_{\rm I})].
\end{align}
This eigenvalue has considerable dependence on wave vector, see Fig.\ \ref{lam-phi}. 
For example, $\lambda_{\phi}^{+}$ becomes the most negative for ${\bf q}_{\rm I}a_{\rm I}=(\pi,0)$, and 
additionally, it will not become positive unless $K_1> 2.947 {\cal D}$, approximately.  This shows that a traveling 
mode along the $x_{\rm I}$ axis destabilizes the {\sc rs}, and $K_1> K_{1,\rm min} \approx 2.947\, {\cal D}$ is required
for stability in the {\sc nn}-model. There is another instability where $\lambda_{\phi}^{-}$ is less than zero
for ${\bf q}_{\rm I}a_{\rm I}=(0,\pi)$ when $K_1$ becomes less than $2.947 {\cal D}$.
Then, the stability requirement for $\lambda_{\theta}^{+}>0$ in
Fig.\ \ref{lam-theta} is satisfied even for $K_3=0$. 

\subsubsection{{\sc nn}-model mode eigenfrequencies}
\begin{figure}
\includegraphics[width=\figwidth,angle=0]{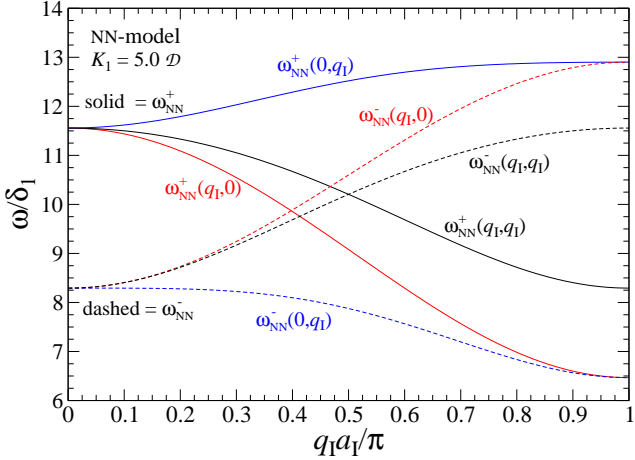}
\caption{\label{wnn-k1d} The mode frequencies $\omega_{\rm NN}^{\pm}$ in units of $\delta_1=\gamma_{\rm e}{\cal D}/\mu$,
in the {\sc nn}-model, for anisotropy $K_1=5\, {\cal D}$, well above that needed for {\sc rs} stability, together with $K_3=0$, for
wave vectors given in the island principal coordinates.}
\end{figure}

For the {\sc nn}-model only, the matrices $\bm{M}_{\theta}$ and $\bm{M}_{\phi}$ have the 
{\em same} eigenvectors, $(\psi^{\pm})^{\dagger}=\frac{1}{\sqrt{2}}(1,\pm 1)$.
A time derivative of equations (\ref{matrix}) separates $\phi$ and $\theta$ solutions, 
\be
\ddot{\psi}_{\phi}=-\left(\frac{\gamma_e}{\mu}\right)^2 \bm{M}_{\theta}\bm{M}_{\phi} \psi_{\phi},
\ee
and similarly for $\ddot{\psi}_{\theta}$. Either eigenvector $\psi_{\phi}=\psi^{+}$
or $\psi_{\phi}=\psi^{-}$ is a separate solution to these equations.
Assuming $e^{-\ii \omega t}$ times dependencies, the eigenfrequencies in the {\sc nn}-model only,
for both $\theta$ and $\phi$ oscillations, \change{which remain intrinsically coupled,} are given by
\be
\omega_{\rm NN}^{\pm}({\bf q}) = \frac{\gamma_e}{\mu}\sqrt{\lambda_{\theta}^{\pm}({\bf q})\lambda_{\phi}^{\pm}({\bf q})},
\ee
\change{
where the natural choice for frequency unit is
\be
\delta_1 \equiv \frac{\gamma_e}{\mu} {\cal D}.
\ee
}
\begin{figure}
\includegraphics[width=\figwidth,angle=0]{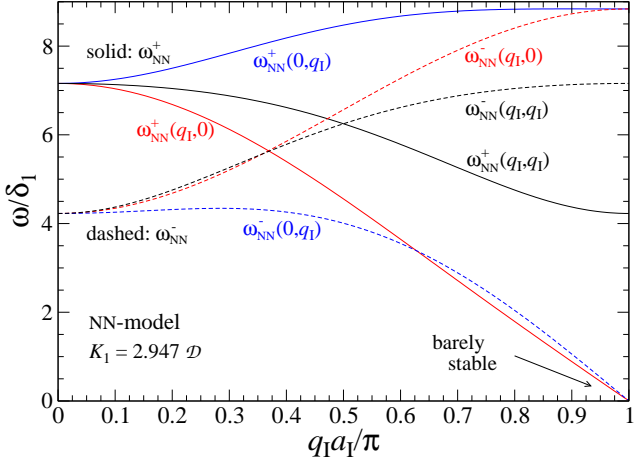}
\caption{\label{wnn-k1min} The mode frequencies $\omega_{\rm NN}^{\pm}$ in units of $\delta_1=\gamma_{\rm e}{\cal D}/\mu$,
in the {\sc nn}-model, for the minimum value, $K_1=K_{1,\rm min} \approx 2.947\, {\cal D}$, that insures {\sc rs} stability, 
together with $K_3=0$, for wave vectors along directions in the island principal coordinates.}
\end{figure}
Some typical mode frequencies are shown in Fig.\ \ref{wnn-k1min}, first for anisotropy at the stability limit, 
$K_1=K_{1,\rm min} \approx 2.947\, {\cal D}$, with $K_3=0$.
The wave vectors are given in the island coordinate system. 
The {\sc rs} mode frequencies in the \change{[10] and [01]} directions are different because the net nonzero
magnetization is along the [10] direction, breaking the symmetry.
The most important feature is the softening of the modes in \change{[10] and [01]} directions towards zero frequency as 
$q_{\rm I}a_{\rm I}\rightarrow \pi$, signifying the {\sc rs} instability.
When the anisotropy is increased to $K_1=5.0\, {\cal D}$, Fig.\ \ref{wnn-k1d}, all modes oscillate well above zero frequency.

While the frequency spectrum might be useful for experimental detection of a state, it is important to go beyond the
{\sc nn}-model and include the modifications due to longer range dipole interactions.

%


\section{Effects of long-range-dipole ({\sc lrd}) interactions}
To get a better description, dipole interactions beyond {\sc nn} must be included.
Consider two islands' dipoles, one at a site {\bf n} and another at displacement {\bf r},
on site ${\bf m=n+r}$.
While the spins are written using their $(x,y,z)$ components, it is best to describe the
displacements in integer {\sc nn} island coordinates, $(x_{\rm I}, y_{\rm I})$, {\it i.e.}
\be
{\bf r} = \left(x_{\rm I}{\hat{\bf x}}_{\rm I}+y_{\rm I}{\hat{\bf y}}_{\rm I}\right)a_{\rm I},
\ee
in contrast to vertex coordinates $(x,y)$, meaning
\be
{\bf r} = \left(x{\hat{\bf x}}+y{\hat{\bf y}}\right)a_{\rm v}.
\ee
The transformation between the two is
\be
x=\tfrac{1}{{2}}\left(x_{\rm I}-y_{\rm I}\right), \quad y=\tfrac{1}{{2}}\left(x_{\rm I}+y_{\rm I}\right).
\ee
When $(x_{\rm I}+y_{\rm I})$ is even (or $x$ and $y$ both integers), the displacement stays on the 
same sublattice (AA or BB bond).
When $(x_{\rm I}+y_{\rm I})$ is odd (or $x$ and $y$ both half-integers), the displacement goes from one 
sublattice to the other (AB bond).
The separation is $r=\sqrt{x_{\rm I}^2+y_{\rm I}^2}\; a_{\rm I}$.
The dipole interaction is reduced from its {\sc nn}-strength ${\cal D}$ by a factor
\be
\rho^3 \equiv r^3/a_{\rm I}^3 = (x_{\rm I}^2+y_{\rm I}^2)^{3/2}.
\ee
From (\ref{Ham}), the island pair dipole interaction is
\begin{align}
\label{Unm}
u_{\bf nm} = & 
  -\frac{\cal D}{\rho^3} \left[ \left(\tfrac{1}{2}-\tfrac{3x_{\rm I}y_{\rm I}}{\rho^2}\right)\mu_{\bf n}^x \mu_{\bf m}^x 
+\left(\tfrac{1}{2}+\tfrac{3x_{\rm I}y_{\rm I}}{\rho^2}\right)\mu_{\bf n}^y \mu_{\bf m}^y \right. \nonumber \\ 
+ & \left. \tfrac{3}{2\rho^2}\left(x_{\rm I}^2-y_{\rm I}^2\right) (\mu_{\bf n}^x \mu_{\bf m}^y+\mu_{\bf n}^y \mu_{\bf m}^x)
-\mu_{\bf n}^z \mu_{\bf m}^z \right].
\end{align}

Consider first an AB bond, where $(x_{\rm I}+y_{\rm I})$ is odd. 
The dipoles are labeled $\bm{\hat{\mu}_{\bf n}}={\bf A}_{\bf n}$ 
and $\bm{\hat{\mu}_{\bf m}}={\bf B}_{\bf m}$.
The dipoles have slight angular deviations from the equilibrium {\sc rs}, as in (\ref{AB-devs}).
The terms needed are expanded up to quadratic order in the deviations, such as
\begin{align}
\label{AxBx}
A_{\bf n}^x B_{\bf m}^x \approx 
\cn \cm & \left[(1-\tfrac{1}{2}\phi_{\bf n}^2-\tfrac{1}{2}\phi_{\bf m}^2
+\phi_{\bf n}\phi_{\bf m}) \cos\phi_A^0 \sin\phi_A^0  \right. \nonumber \\
& \left. -\phi_{\bf n}\sin^2\!\phi_A^0-\phi_{\bf m} \cos^2\!\phi_A^0 \right], 
\end{align}
where the out-of-plane deviation factor is
\be
\cn \cm  \approx 1-\tfrac{1}{2}\theta_{\bf n}^2 -\tfrac{1}{2}\theta_{\bf m}^2.
\ee
There is a similar term for $y$ components,
\begin{align}
\label{AyBy}
A_{\bf n}^y B_{\bf m}^y \approx 
\cn \cm & \left[(1-\tfrac{1}{2}\phi_{\bf n}^2-\tfrac{1}{2}\phi_{\bf m}^2
+\phi_{\bf n}\phi_{\bf m}) \cos\phi_A^0 \sin\phi_A^0 \right. \nonumber \\
& \left. +\phi_{\bf n}\cos^2\!\phi_A^0+\phi_{\bf m} \sin^2\!\phi_A^0 \right]. 
\end{align}
Finally, there is the cross coupling of components, 
\be
\label{AxBy}
A_{\bf n}^x B_{\bf m}^y+A_{\bf n}^y B_{\bf m}^x \approx 
\cn\cm (1-\tfrac{1}{2}\phi_{\bf n}^2-\tfrac{1}{2}\phi_{\bf m}^2-\phi_{\bf n}\phi_{\bf m}).
\ee
One can see that the pair's dipolar energy $u_{\rm AB}$ contains zeroth order, first order, and
quadratic order terms.  

Consider instead an AA pair with energy $u_{\rm AA}$, or similarly, a BB pair, where $(x_{\rm I}+y_{\rm I})$ 
is even.  For $u_{\rm AA}$ the needed expressions are 
\begin{align}
\label{AxAx}
A_{\bf n}^x A_{\bf m}^x \approx 
\cn \cm & \left[(1-\tfrac{1}{2}\phi_{\bf n}^2-\tfrac{1}{2}\phi_{\bf m}^2) \cos^2\!\phi_A^0
+ \phi_{\bf n}\phi_{\bf m} \sin^2\!\phi_A^0  \right. \nonumber \\
& \left. -(\phi_{\bf n}+\phi_{\bf m})\sin\phi_A^0\cos\phi_A^0  \right], 
\end{align}
\begin{align}
A_{\bf n}^y A_{\bf m}^y \approx 
\cn \cm & \left[(1-\tfrac{1}{2}\phi_{\bf n}^2-\tfrac{1}{2}\phi_{\bf m}^2) \sin^2\!\phi_A^0
+ \phi_{\bf n}\phi_{\bf m} \cos^2\!\phi_A^0  \right. \nonumber \\
& \left. +(\phi_{\bf n}+\phi_{\bf m})\sin\phi_A^0\cos\phi_A^0  \right], 
\end{align}
\begin{align}
\label{AxAy}
A_{\bf n}^x A_{\bf m}^y & +A_{\bf n}^y A_{\bf m}^x \approx \nonumber \\
& \cn\cm \left[(1-\tfrac{1}{2}\phi_{\bf n}^2-\tfrac{1}{2}\phi_{\bf m}^2-\phi_{\bf n}\phi_{\bf m})
\sin 2\phi_A^0  \right.
\nonumber \\
& \left. +(\phi_{\bf n}+\phi_{\bf m}) \cos 2\phi_A^0 \right].
\end{align}
For $u_{\rm BB}$ because the equilibrium directions are different on the B-sublattice, the expressions
are also different, swapping the factors of $\sin\phi_A^0$ and $\cos\phi_A^0$,
\begin{align}
\label{BxBx}
B_{\bf n}^x B_{\bf m}^x \approx 
\cn \cm & \left[(1-\tfrac{1}{2}\phi_{\bf n}^2-\tfrac{1}{2}\phi_{\bf m}^2) \sin^2\!\phi_A^0
+ \phi_{\bf n}\phi_{\bf m} \cos^2\!\phi_A^0  \right. \nonumber \\
& \left. -(\phi_{\bf n}+\phi_{\bf m})\sin\phi_A^0\cos\phi_A^0  \right], 
\end{align}
\begin{align}
B_{\bf n}^y B_{\bf m}^y \approx 
\cn \cm & \left[(1-\tfrac{1}{2}\phi_{\bf n}^2-\tfrac{1}{2}\phi_{\bf m}^2) \cos^2\!\phi_A^0
+ \phi_{\bf n}\phi_{\bf m} \sin^2\!\phi_A^0  \right. \nonumber \\
& \left. +(\phi_{\bf n}+\phi_{\bf m})\sin\phi_A^0\cos\phi_A^0  \right], 
\end{align}
\begin{align}
\label{BxBy}
B_{\bf n}^x B_{\bf m}^y & +B_{\bf n}^y B_{\bf m}^x \approx \nonumber \\
& \cn\cm \left[(1-\tfrac{1}{2}\phi_{\bf n}^2-\tfrac{1}{2}\phi_{\bf m}^2-\phi_{\bf n}\phi_{\bf m})
\sin 2\phi_A^0  \right.
\nonumber \\
& \left. -(\phi_{\bf n}+\phi_{\bf m}) \cos 2\phi_A^0 \right].
\end{align}

%
\subsection{The shifted equilibrium}
%
\begin{figure}
\includegraphics[width=\figwidth,angle=0]{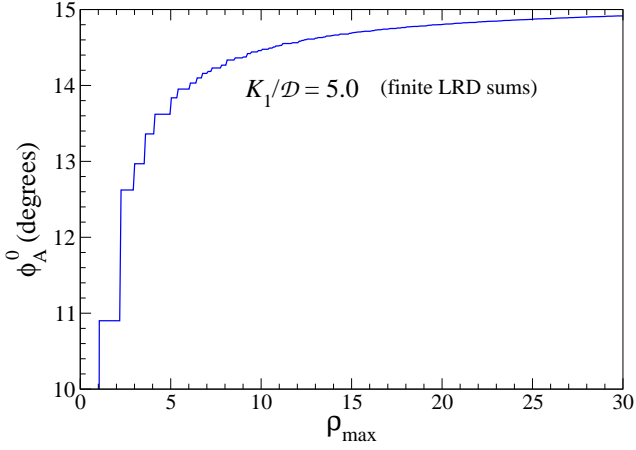}
\caption{\label{phiA-v-rho} \change{The effect of adding longer-range dipole interactions on the remanent 
state's equlibrium tilting angle $\phi_A^0$, Eq.\ (\ref{lreq}), as a function of the largest radius used
in the needed sum $s_{\rm AB}$ of equation (\ref{sums}). The first large jump at $\rho_{\rm max}=\sqrt{5}$
is due to fourth nearest neighbors.}}
\end{figure}
\label{lrdeq}
Including {\sc lrd} interactions, the Hamiltonian can be expressed 
\be
H=H^{(0)}+H^{(1)}+H^{(2)},
\ee
where the superscripts indicates the zeroth, linear, and quadratic terms in the deviations
around equilibrium.
The equilibrium still has opposing in-plane tilting, $\phi_B^0 = -\phi_A^0$.
The terms in $H$ come partly from AB bonds and partly from 
AA bonds (nearly equivalent to BB bonds), as well as the anisotropy. 
At the equilibrium, the linear part $H^{(1)}$ vanishes.
There are no linear terms in $\theta_{\bf n}$ in $H$, implying that the equilibrium
still has values $\theta_{\bf n}=0$ for all sites, and thus all $s_{\bf n}=0,\ c_{\bf n}=1$. 
The zeroth order terms, not containing $\theta_{\bf n}$ nor $\phi_{\bf n}$, are
very simple in Eqs.\ (\ref{Unm}) through (\ref{BxBy}) and are easy to apply to obtain
$H^{(0)}$, employing inversion symmetry of the system, and dividing by two to avoid 
double counting.
The equilibrium energy of an island interacting with the entire system via
{\sc lrd} interactions, per island, $\varepsilon_{\infty}=H^{(0)}/N$, is found to be
\begin{align}
\label{H0}
\varepsilon_{\infty} &=  K_1 \sin^2\!\phi_A^0-\frac{\cal D}{4}\left(\sum_{x_{\rm I},y_{\rm I}}^{\rm AB} \frac{ \sin 2\phi_A^0}{\rho^3}
 +\sum_{x_{\rm I},y_{\rm I}}^{\rm AA} \frac{1}{\rho^3}\right).
\end{align}
The first sum is over AB bonds, where $\rho^2=1,5,9,13,$ etc.
The second sum is over AA bonds, with $\rho^2=2,4,8,$ etc. 
Estimates of the sums are
\begin{align}
\label{sums}
s_{\rm AB} & = \sum_{x_{\rm I},y_{\rm I}}^{\rm AB} \frac{1}{\rho^3} \approx 5.8397, \quad
s_{\rm AA} = \sum_{x_{\rm I},y_{\rm I}}^{\rm AA} \frac{1}{\rho^3} \approx 3.1926.
\end{align}
Then with infinite-range dipole interactions, the energy per island is 
\be
\label{H0s}
\varepsilon_{\infty} =  K_1 \sin^2\!\phi_A^0 - \tfrac{1}{4}{\cal D}\left(s_{\rm AB}\sin 2\phi_A^0 +s_{\rm AA}\right).
\ee
The equilibrium tilting of the dipoles takes place at the minimum of $H^{(0)}$, which gives
\be
\label{lreq}
\tan 2\phi_A^0 =  -\tan 2\phi_B^0 = \frac{s_{\rm AB} \cal D}{2K_1} \approx \frac{2.9198 \cal D}{K_1}.
\ee
Thus, the effect of infinite-range dipolar interactions is to {\em increase} the inward
tilting of the two sublattices towards each other, compared to the {\sc nn}-model [Eq.\ (\ref{RS0})],
see Fig.\ \ref{eps+phiA}. 
\change{The change in tilting as longer-range dipolar interactions are included is shown Fig.\ 
\ref{phiA-v-rho}, as a function of the maximum neighbor distance $\rho_{\rm max}$ used in the sum 
$s_{\rm AB}$, for $K_1/{\cal D}=5$.  The largest jump (beyond {\sc nn} interactions) occurs at 
$\rho_{\rm max}=\sqrt{5}$, where $\phi_A^0$ changes from $10.9^{\circ}$ to $12.6^{\circ}$.}
This is attributed to {\em fourth} nearest neighbor interactions (eight AB bonds with
$\rho^2=5$) that try to align the A and B lattices.
The AA bonds do not shift the equilibrium, as the sum $s_{\rm AA}$ plays
no role in the formula for $\phi_A^0$, but they contribute to the energy.

\subsection{Dynamics with long-range-dipole interactions} 
The last term in the Hamiltonian, $H^{(2)}=H_{\phi}+H_{\theta}$, is quadratic in the small deviations 
$\phi_{\bf n}$ and $\theta_{\bf n}$, and controls the dynamics, as in Eqs.\ (\ref{dots}) and (\ref{matrix}).
Equations (\ref{Unm}) through (\ref{BxBy}) give the contributions of arbitrary range dipole interactions to 
$H^{(2)}$, and implicitly define the matrices $\bm{M}_{\phi}$ and $\bm{M}_{\theta}$.
Once {\sc lrd} interactions are included, the matrices $\bm{M}_{\phi}$ and $\bm{M}_{\theta}$
do not have the same eigenvectors, so a more general procedure is needed to get the dynamic
modes.

The dynamics in (\ref{matrix}) is still solved using traveling waves written in the {\sc nn} island coordinates. 
While locating an island by ${\bf n}=n_{x_{\rm I}} {\bf x}_{\rm I}+n_{y_{\rm I}}{\bf y}_{\rm I}$, 
a displacement to another island of a dipole pair is expressed as 
${\bf r} = x_{\rm I}{\bf x}_{\rm I}+y_{\rm I}{\bf y}_{\rm I}$.
Assume waves on both sublattices varying in time as $e^{-\ii \omega t}$ (suppressed in the formulas), 
the same as in the {\sc nn}-model,
\begin{align}
\theta_{\bf n}^{\rm A} &= a_{\theta} e^{\ii {\bf q}\cdot {\bf n}}, \quad 
\phi_{\bf n}^{\rm A} = a_{\phi} e^{\ii {\bf q}\cdot {\bf n}},  \quad \text{A-sites}, \nonumber \\
\theta_{\bf n}^{\rm B} &= b_{\theta} e^{\ii {\bf q}\cdot {\bf n}}, \quad 
\phi_{\bf n}^{\rm B} = b_{\phi} e^{\ii {\bf q}\cdot {\bf n}},  \quad \text{B-sites}. 
\end{align}
The allowed wave vectors were given in (\ref{qvec}).

The matrix form of the dynamic equations (\ref{matrix}) 
involves sums over matrix elements with the spin field components,
\bn
\label{lrdyn}
-\ii \omega \phi_{\bf n} &=&  \tfrac{\gamma_{\rm e}}{\mu}
(M_{\theta,{\bf n,n}}\; \theta_{\bf n} +\sum_{{\bf r}\ne 0} M_{\theta,{\bf n,n+r}}\; \theta_{\bf n+r}),
\nonumber \\
\ii \omega \theta_{\bf n} &=&  \tfrac{\gamma_{\rm e}}{\mu}
( M_{\phi,{\bf n,n}}\; \phi_{\bf n} +\sum_{{\bf r}\ne 0} M_{\phi,{\bf n,n+r}}\; \phi_{\bf n+r}).
\en
This pair of equations becomes four equations when both A and B sublattices are considered.
The matrix elements are either within a sublattice ($M_{\bf n,m}^{\rm AA}, M_{\bf n,m}^{\rm BB}$)
or between sublattices ($M_{\bf n,m}^{\rm AB}, M_{\bf n,m}^{\rm BA}$). They are derived from
the quadratic terms in Eqs.\ (\ref{Unm}) through (\ref{BxBy}).  For example, the coefficient 
of $\phi_{\bf n}^2$  in $A_{\bf n}^x B_{\bf m}^x$ in (\ref{AxBx}) contributes to $M_{\phi,\bf n,n}^{\rm AB}$, 
while the coefficient of $\phi_{\bf n}\phi_{\bf m}$ contributes to $M_{\phi,\bf n,m}^{\rm AB}$.
 
With the wave assumption, the equations comprise coupled systems with $2\times 2$ matrices named $\bm{m}$
and $\bm{n}$ for compact notation,
\be
-\ii \omega \left[ \begin{array}{c} a_{\phi} \\ b_{\phi} \end{array} \right]
=\left[\begin{array}{cc} m_{aa} & m_{ab} \\ m_{ba} & m_{bb} \end{array} \right]
            \left[ \begin{array}{c} a_{\theta} \\ b_{\theta} \end{array} \right],
\ee
\be
-\ii \omega \left[ \begin{array}{c} a_{\theta} \\ b_{\theta} \end{array} \right]
=-\left[\begin{array}{cc} n_{aa} & n_{ab} \\ n_{ba} & n_{bb} \end{array} \right]
            \left[ \begin{array}{c} a_{\phi} \\ b_{\phi} \end{array} \right].
\ee
This is identical to a single $4\times 4$ eigenvalue problem, 
\be
\label{4x4}
\left[\begin{array}{cc} \begin{array}{cc} 0 & 0 \\ 0 & 0 \end{array} & \bm{m} \\
 \bm{n} & \begin{array}{cc} 0 & 0 \\ 0 & 0 \end{array} \end{array} \right]
\left[ \begin{array}{c} a_{\phi} \\ b_{\phi} \\ \ii a_{\theta} \\ \ii b_{\theta} \end{array} \right]
=  \omega \left[ \begin{array}{c} a_{\phi} \\ b_{\phi} \\ \ii a_{\theta} \\ \ii b_{\theta} \end{array} \right].
\ee
Symbolically, the $2\times 2$ matrices off the diagonal are 
\be
\bm{m} = \frac{\gamma_{\rm e}}{\mu} \bm{M}_{\theta}, \quad 
\bm{n} = \frac{\gamma_{\rm e}}{\mu} \bm{M}_{\phi},
\ee
as projected onto the traveling wave solutions.
More specifically, the matrix elements of $\bm{m}$ and $\bm{n}$ are given by sums of the interactions 
as derived from $H^{(2)}$. For instance, $m_{aa}$ comes from the AA dipole interactions as well as
the on-site anisotropy. Supposing {\bf n} is an  A-site, with {\bf n+r} also an A-site,
\be
m_{aa}=\frac{\gamma_{\rm e}}{\mu} 
\left(M_{\theta, \bf n,n}+\sum_{\bf r}^{\rm AA} M_{\theta, \bf n, n+r} e^{\ii {\bf q}\cdot {\bf r}}\right).
\ee
The symbol $\sum_{\bf r}^{\rm AA}$ indicates summing over displacements on one sublattice ($x_{\rm I}+y_{\rm I}=$even).
The same expression gives $m_{bb}=m_{aa}$.  For the AB couplings there is also symmetry,
\be
m_{ab}=m_{ba} = \frac{\gamma_{\rm e}}{\mu}
 \sum_{\bf r}^{\rm AB} M_{\theta,\bf n,n+r} e^{\ii {\bf q}\cdot {\bf r}}
\ee
where $\sum_{\bf r}^{\rm AB}$ indicates summing over displacements from one sublattice to the other 
($x_{\rm I}+y_{\rm I}=$ odd). 
The full {\bf q}-dependence of the matrix elements of $\bm{m}$ and $\bm{n}$ is shown in Appendix \ref{elements}.

\subsubsection{Stability requirements with {\sc lrd} interactions}
%
\begin{figure}
\includegraphics[width=\figwidth,angle=0]{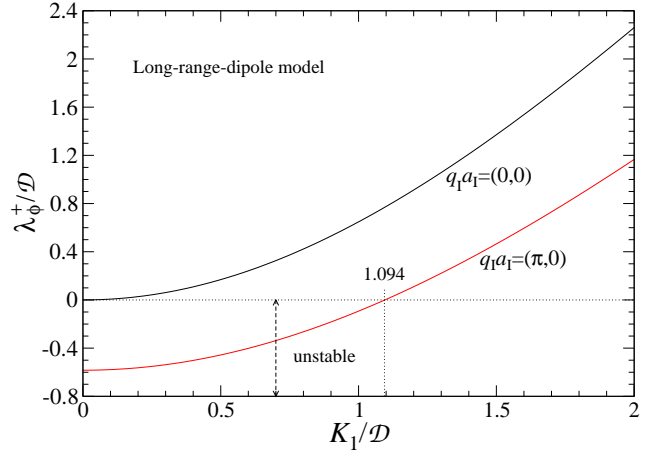}
\caption{\label{lam-phi-lrd}
With infinite-range dipole interactions, the in-plane energy eigenvalue $\lambda_{\phi}^{+}$ vs.\ in-plane anisotropy, for 
two wave vectors \change{in island coordinates}, showing that stability requires $K_1>\approx 1.094\, {\cal D}$, significantly less than 
$K_1 > 2.947\, {\cal D}$ in the {\sc nn}-model.}
\end{figure}
Similar to the {\sc nn}-model, the eigenvalues of matrices $\bm{n}$ and $\bm{m}$ must be positive for stability of the 
{\sc rs} when {\sc lrd} interactions are included. The controlling requirement is due to the eigenvalues 
$\lambda_{\phi}^{\pm}$ of matrix $\bm{n}$.  A general formula for its eigenvalues (as for any $2\times 2$ matrix) is
\be
\lambda_{\phi}^{\pm} = \tfrac{1}{2}(n_{aa}+n_{bb})\pm \sqrt{\tfrac{1}{4}(n_{aa}+n_{bb})^2-(n_{aa}n_{bb}-n_{ab}n_{ba})}.
\ee
The most unstable eigenvalue occurs at ${\bf q}a_{\rm I}=(\pi,0)$, where the sums needed (see Appendix \ref{elements}) become
\begin{align}
f_{\rm odd}(\pi,0) &= d_{\rm evn}(\pi,0)= f_{\rm evn}^{xy}(\pi,0)=0, \\
f_{\rm evn}(\pi,0) &\approx -0.93546,\ d_{\rm odd}(\pi,0)\approx -3.71107
\end{align}
Then the eigenvalues are $\lambda_{\phi}^{\pm} = n_{aa}\pm n_{ab}$, or
\begin{align}
\lambda_{\phi}^{\pm}(\pi,0) & = 2K_1  \cos 2\phi_A^0 +\tfrac{1}{2}{\cal D}\left[ s_{\rm AB} \sin 2\phi_A^0 \right. \nonumber \\
& \left. + s_{\rm AA} -f_{\rm evn}(\pi,0) \pm 3\, d_{\rm odd}(\pi,0)  \right].
\end{align}
Because $d_{\rm odd}(\pi,0)$ is negative, the eigenvalue $\lambda_{\phi}^{+}(\pi,0)$ is smallest. It 
is responsible for an instability at $K_1<1.094\; {\cal D}$, see Fig.\ \ref{lam-phi-lrd}.
As a result, {\sc lrd} interactions enhance the stability of the remanent state, meaning that even rather weak
uniaxial anisotropy of the islands will be able to maintain that state.

\subsubsection{Mode frequencies with {\sc lrd} interactions}
It is shown in Appendix \ref{eigen} that the eigenfrequencies of the $4\times 4$ eigenvalue problem (\ref{4x4}) are 
obtained from
\be
\label{Awe0}
\omega^2 = \frac{1}{2}\left[\left(\bm{m}^{T}\cdot \bm{n}\right)
\pm \sqrt{\left(\bm{m}^{T}\cdot \bm{n}\right)^2-4|\bm{m}| |\bm{n}|}\; \right].
\ee
The two solutions can be labeled as $\omega_{\infty}^{\pm}({\bf q})$, where $\infty$ indicates keeping 
{\sc lrd} interactions to unlimited distances.  The formula contains a dot product of the two $2\times 2$
dynamic matrices, as well as a product of their determinants. It should apply to any spin dynamics
problem involving a two-sublattice partitioning of the system.  The two modes might be considered as 
acoustic and optic modes, however, such identification really depends on the spin components being 
considered.  

Following the procedure in Appendix \ref{eigen}, and applying Eq.\ \ref{Awe0}, the dispersion relations
for modes of excitation from a remanent state were obtained for various cases of anisotropy.

\subsubsection{Changes in dynamics -- second nearest neighbors}
%
\begin{figure}
\includegraphics[width=\figwidth,angle=0]{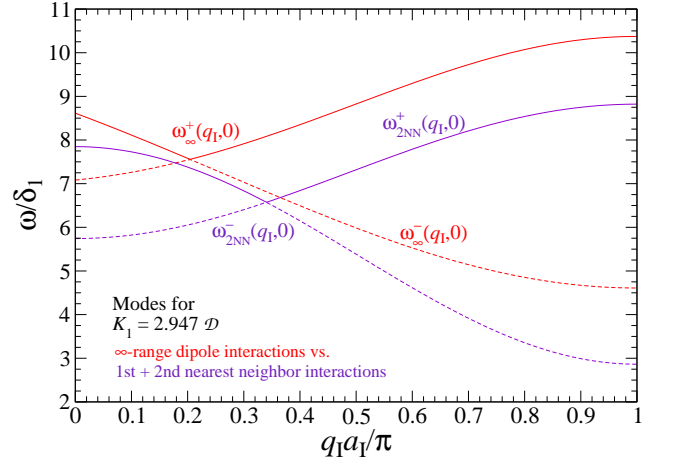}
\caption{\label{w10-2inf-k1a}  Comparison of the \change{[10]} {\sc rs} dispersion relations 
\change{(wave vectors in island coordinates)} using the 1$^{\rm st}$+2$^{\rm nd}$ neighbors model 
($\omega_{2\rm NN}^{\pm}$, indigo) and the model with all $\infty$-range dipole interactions 
($\omega_{\infty}^{\pm}$, red) calculated using sums out to $\rho\approx 4000$, at 
$K_1 = 2.974\, {\cal D}$, just above the minimum needed for stability in the {\sc nn}-model. The frequencies are 
higher with more dipole terms and the crossing point shifts to lower wave vector.}
\end{figure}
To see the general trend due to going beyond {\sc nn}-interactions, first we include only the interactions
up to second nearest neighbors ({\sc 2nn}), with displacements $(x,y) = (\pm 1,0)a_{\rm v}, (0,\pm 1)a_{\rm v}$, which 
are the nearest AA or BB bonds. 
It is straightforward to show that the equilibrium angles $\phi_A^0$ are the same as in the {\sc nn}-model.
The procedure in Eq.\ \ref{Awe0} gives the eigenfrequencies, using sums truncated at 2$^{\rm nd}$-nearest neighbors.
A partial mode spectrum is shown in Fig. \ref{w10-2inf-k1a}, for $K_1=2.947 {\cal D}$, where the {\sc 2nn}-model is compared
to the model using infinite-range dipole interactions.
The {\sc 2nn} interactions already lift eigenfrequencies enough to relieve the instability that is present in the
{\sc nn}-model, compare Fig.\ \ref{wnn-k1min}. 
Once {\sc lrd}-interactions to infinite range are included, the frequencies are raised further, and the notable
crossing point between higher and lower modes shifts to lower frequency.
Surprisingly, it is not an avoided crossing.

\subsubsection{Changes in dynamics -- infinite range dipole interactions}
%
\begin{figure}
\includegraphics[width=\figwidth,angle=0]{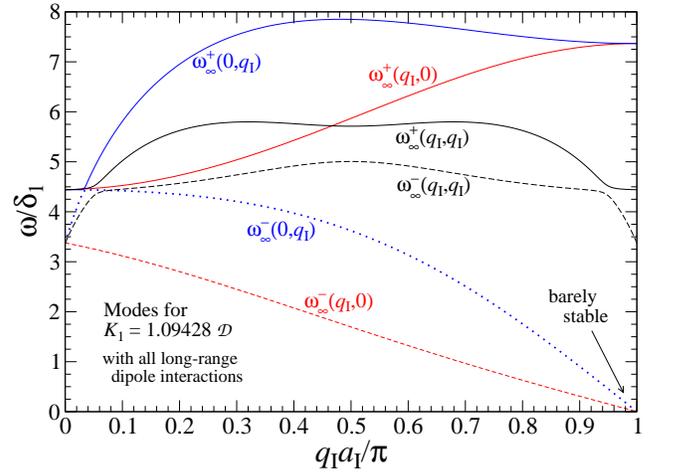}
\caption{\label{w-lrd-k1c}  Dispersion relations \change{in island coordinates} using all dipole interactions, 
for a remanent state with $K_1 = 1.09428\, {\cal D}$ and $K_3=0$, just above the minimum needed for stability, 
calculated using sums out to $\rho\approx 10000$.  Both the \change{[10] and [01]} $\omega_{\infty}^{-}$ dispersion 
relations go unstable at $q_{\rm I}a_{\rm I}=\pi$ for smaller $K_1$.}  
\end{figure}
The stability limit at $K_1 \approx 1.094\, {\cal D}, K_3=0,$ with infinite-ranged dipole interactions can be verified 
by finding the mode dispersion relations for wave vectors along \change{[10], [01], and [11]} directions in island coordinates, 
see Fig.\ \ref{w-lrd-k1c}.  
It is found that the $\omega_{\infty}^{-}$ modes along both the \change{[10] and [01]} directions go to zero for $q_{\rm I}a_{\rm I}=\pi$,
similar to the behavior in the {\sc nn}-model, compare Fig.\ \ref{wnn-k1min}, even though the [01] frequency is higher, away from
the stability limit point.
One concludes that spin waves traveling either parallel to \change{(along [10])} or perpendicular to \change{(along [01])} the magnetic
moment of the {\sc rs} contribute to its instability at weak uniaxial anisotropy in the islands.
Even so, very little uniaxial anisotropy is needed to stabilize a {\sc rs} under the influence of long-range dipole interactions.

\begin{figure}
\includegraphics[width=\figwidth,angle=0]{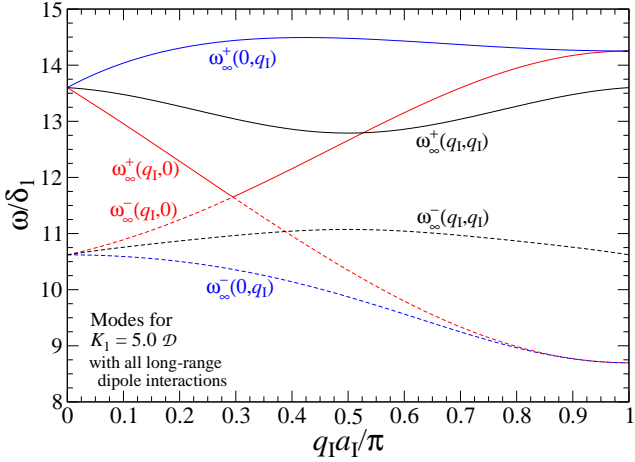}
\caption{\label{w-lrd-k1d}  Dispersion relations using all dipole interactions for a {\sc rs} at 
$K_1 = 5.0\, {\cal D}$,  calculated using sums out to $\rho\approx 4000$.
While the $\omega_{\infty}^{-}$ and $\omega_{\infty}^{+}$ frequencies along \change{[01] and [11] island directions} are far from each other,
they touch or /cross at a point in the \change{[10]} direction, i.e., for wave vectors parallel to the {\sc rs} magnetization.}
\end{figure}
\begin{figure}
\includegraphics[width=\figwidth,angle=0]{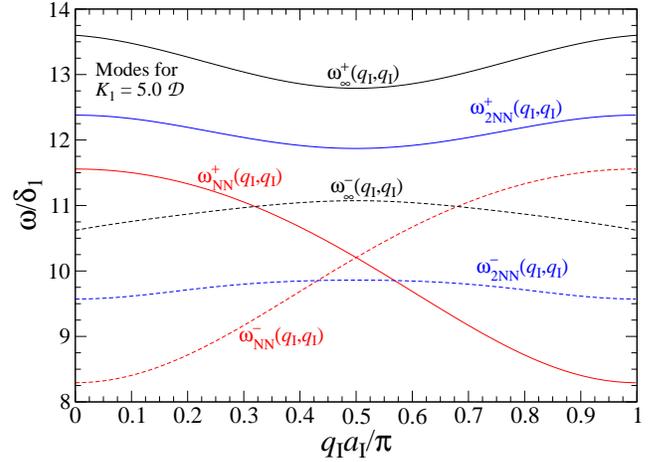}
\caption{\label{w11-k1d}  \change{Dispersion relations at $K_1 = 5.0\, {\cal D}, K_3=0$ for wave vectors along the [11]
island direction only, comparing the calculations of the {\sc nn} model ($\omega_{\rm NN}^{\pm}$) , the model with 1$^{\rm st}$ 
and 2$^{\rm nd}$ nearest neighbors ($\omega_{\rm 2NN}^{\pm}$), and the model with all {\sc lrd} interactions 
($\omega_{\infty}^{\pm}$).}} 
\end{figure}

Another example of the dispersion relations is shown in Fig.\ \ref{w-lrd-k1d} for $K_1=5\, {\cal D}, K_3=0$.
The shapes have changed noticeably from how they appeared at the stability limit for $K_1$. 
Note that the frequencies along \change{[10]} touch or cross now at $q_{\rm I}a_{\rm I}\approx 0.3\pi$, 
while those along \change{[01] and [11]} remain very highly separated for all $q_{\rm I}$.
This is in strong contrast to the result in Fig.\ \ref{wnn-k1d} for the {\sc nn}-model above the stability limit.
%
The conclusion is that dipolar interactions are especially influential along the \change{[11] island} direction in keeping
the higher and lower mode frequencies separated.
\change{This effect is highlighted in Fig.\ \ref{w11-k1d}, where the frequencies along [11] are compared for the
{\sc nn} model ($\omega_{\rm NN}^{\pm}$), the model with 1$^{\rm st}$ and 2$^{\rm nd}$ nearest neighbors ($\omega_{\rm 2NN}^{\pm}$),
and the model with all {\sc lrd} interactions ($\omega_{\infty}^{\pm}$). A drastic change occurs when the 2$^{\rm nd}$ nearest neighbors
are added, while going from there to very long range interactions shifts the frequencies about 10\% higher.}

Consider next a real square spin-ice material such as that using Permalloy studied by Wang {\it et al.} \cite{Wang06}, with
elongated islands of approximate dimensions 220 nm $\times$ 80 nm $\times$ 25 nm thick.  
Based on the saturation magnetization $M_s=860$ kA m$^{-1}$ multiplying the volume of elliptical islands, the
island magnetic dipole moment was estimated as $\mu=2.97 \times 10^{-16}$ A m$^2$, see Ref.\ \onlinecite{Wysin+13}.
Then supposing a square ice with a close vertex spacing, $a_v=320$ nm (island {\sc nn} spacing $a_{\rm I}=a_v/\sqrt{2}$),  
the {\sc nn} dipolar coupling constant in Eq.\ (\ref{Dd}) is ${\cal D}=7.61\times 10^{-19}$ J. 
Simulations in Ref.\ \onlinecite{Wysin+12} can be used to estimate the anisotropy constants, for the chosen island 
aspect ratios, and they were found \cite{Wysin+13} to be $K_1 = 2.9 \times 10^{-17}$ J and $K_3=6.4 \times 10^{-17}$ J.
These are high compared to room temperature thermal energy, which insures stabilization
of a remanent or other discrete spin-ice state, and truly forces the oscillations to be of small amplitude.
Then the scaled anisotropy parameters needed here are estimated as $K_1=38\; {\cal D}$ and $K_3=84\; {\cal D}$, 
based on this particular geometry of the island lattice.  

\begin{figure}
\includegraphics[width=\figwidth,angle=0]{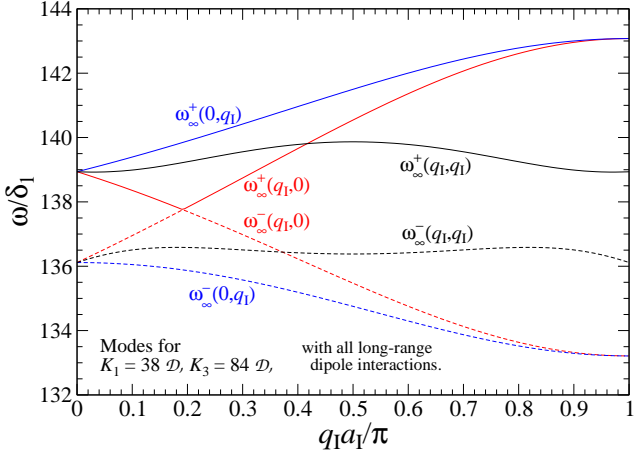}
\caption{\label{w-lrd-kwang}  Dispersion relations \change{in island coordinates} using all dipole interactions, for a remanent state at
realistic spin-ice parameters appropriate to the square spin-ice of Wang {\it et al.} \cite{Wang06}.  
Calculated using sums out to $\rho\approx 4000$. Note the linear dispersions on both sides of the \change{[10]} touching 
point, while the modes stay fairly widely separated along \change{[01] and [11]}.
Further, the modes along \change{[10] and [01]} approach $q_{\rm I}a_{\rm I}\to \pi$ with zero slope.}
\end{figure}

For this realistic model, mode frequencies or wave vectors along \change{[10], [01] and [11]} directions in island 
coordinates are plotted in Fig.\ \ref{w-lrd-kwang}.  
Relative to the examples with weaker anisotropy, the whole spectrum has been shifted to higher frequencies
due to the strong anisotropy in typical spin-ice with greatly elongated islands.
Note again the vivid linearity of dispersion relations along the \change{[10]} direction near $q_{\rm I}=0$,
and in the region near the point where higher and lower frequencies touch.
The modes along \change{[01]}, to the contrary, remain widely separated for the whole range of wave vectors.
Along \change{[11]}, both modes $\omega_{\infty}^{+}$ and $\omega_{\infty}^{-}$ are almost independent of $q_{\rm I}$, 
with only slight variations.
The modes along \change{[10] and [01]} both approach $q_{\rm I}a_{\rm I}\to \pi$ with zero slope.

\section{Conclusions}
An effective model with Heisenberg-like island dipoles influenced by anisotropies and dipole-dipole interactions 
has been applied to find the remanent state properties for square spin ice, including the spin configuration, its energy, 
angular deviation eigenvalues, and the normal modes of oscillation about a {\sc rs}.
The model allows the net dipole of each island to deviate continuously in direction from its long axis, while paying 
a cost in anisotropy energy.
This analysis would not be possible if the island dipoles were represented as Ising spins.

The model with {\sc nn}-dipole-interactions, although somewhat limited, was used to describe the static and 
dynamic calculations and to estimate the basic properties.
The {\sc rs} energy (Fig.\ \ref{eps+phiA}) for $K_1>K_{1,\rm min}\approx 2.947\; {\cal D}$ is 
$\varepsilon_{\rm NN}\approx -0.2\; {\cal D}$, compared to the ground state energy $-3{\cal D}$ in the {\sc nn}-approximation. 
Hence, the {\sc rs} exhibits a metastable property.
It has considerably higher energy than a ground state but nevertheless is stabilized from small oscillations by relatively
weak uniaxial anisotropy of the islands, even when including only {\sc nn}-dipole-interactions. 

In the {\sc nn}-model, a {\sc rs} is stable in the absence of planar anisotropy ($K_3=0$) as long as the
uniaxial anisotropy of an island surpasses $K_{1,\rm min}=2.947\; {\cal D}$, where ${\cal D}$ is the {\sc nn} dipolar interaction
strength. 
The instability for weak uniaxial anisotropy ($K_1<K_{1,\rm min}$) can be attributed to in-plane
deviation eigenvalues becoming zero at the limiting anisotropy: $\lambda_{\phi}^{+}(\pi,0)\rightarrow 0$
and $\lambda_{\phi}^{-}(0,\pi)\rightarrow 0$.
The net magnetic moment of the system in the selected remanent state is along the \change{[10] direction of the island lattice
([11] direction of the $xy$ coordinates of the vertex lattice).}
Although the nonzero magnetization {\bf M} breaks the symmetry of the system, \change{modes along island directions
[10] and [01]} both go unstable at $K_{1,\rm min}$, see Fig.\ \ref{wnn-k1min}.
The eigenvalues $\lambda_{\phi}^{\pm}$ become imaginary for $K_1< K_{1,\rm min}$, implying that large
in-plane fluctuations will grow with time for the unstable {\sc rs}.
The out-of-plane deviation eigenvalues $\lambda_{\theta}^{\pm}$ remain positive and do not play a role
in the instability, even for planar anisotropy $K_3=0$, as long as $K_1>K_{1,\rm min}$.

%

A procedure was developed here to include all dipole-dipole interactions of a central site with neighbors at 
any distance on the square lattice.
With dipole-dipole bonds classified as AA or BB (intrasublattice) and AB (intersublattice), it is found that the
AA and BB bonds do not change the {\sc rs} spin angles, but they do affect the dynamic frequencies.
{\sc lrd} interactions cause the sublattice spins to tilt more closely towards each other \change{(closer to the [10] direction)}
compared to their directions in the {\sc nn}-model, see Fig.\ \ref{eps+phiA}.
That extra tilting puts the dipoles into an energetically more favorable configuration for dipole-dipole interactions,
and lowers the {\sc rs} energy, while the state remains metastable.

With infinite-range dipole interactions, the instability of the {\sc rs} for $q_{\rm I}a_{\rm I}\to \pi$ still 
takes place for wave vectors along both the \change{[10] and [01] island directions} for $K_1 < K_{1,\rm min}$, however,
the limiting value decreases to $K_{1,\rm min} = 1.09428\; {\cal D}$.
This implies that the extra dipole interactions beyond {\sc nn} tend to keep the island spins more strongly along the
island axes, with less need for uniaxial anisotropy. 
A remanent state of square spin ice will not be stable for $K_3=0$ unless $K_1 > 1.09428\; {\cal D}$, 
a rather weak anisotropy constraint.
For the model of a realistic square spin ice, the large anisotropy values $K_1/{\cal D}=38$, $K_3/{\cal D}=84$  
very strongly stabilize a remanent state.

The dynamics with {\sc lrd} interactions is determined by a pair of coupled $2\times 2$ eigenvalue problems, equivalent to a
single $4\times 4$ system.
Due to the symmetry properties of the involved matrices, the exact frequency eigenvalues of the $4\times 4$ system can
be calculated.
Generally, the mode frequencies increase as longer range dipole interactions are included.
For realistic parameters for square spin-ice, Fig.\  \ref{w-lrd-kwang}, the mode frequencies are fairly high
already due to the islands' anisotropies.
The modes $\omega_{\infty}^{\pm}$ with wave vectors along \change{[10]} (parallel to the {\sc rs} magnetization) touch at
one point and display a linear behavior at lower wave vectors.  
The other modes $\omega_{\infty}^{\pm}$ with wave vectors along \change{[01] and [11]} stay well separated. 
Along \change{[11]} that mode separation appears to be due to the {\sc lrd} interactions.
These calculations are expected to be applicable for finding state stability and mode properties in other
distinct states of spin ice and can be adapted to different lattices.

\appendix
\section{Full {\sc lrd} Hamiltonian $H^{(2)}$}
Here the quadratic terms in expression (\ref{Unm}) for dipole pair energy are fully 
expanded in the small deviations $\theta_{\bf n}$ and $\phi_{\bf n}$ 
and a complete expression for $H^{(2)}$ that determines the dynamics is given.
The quadratic part of an AB pair interaction energy can be written as
\be
u_{\rm AB}^{(2)} = u_{\rm AB}^{xx}+u_{\rm AB}^{yy}+u_{\rm AB}^{xy}+u_{\rm AB}^{zz}.
\ee
Eqs.\ (\ref{AxBx}) through (\ref{AxBy}) are used to get the contributions to $u_{\rm AB}^{(2)}$,
\begin{align}
u_{\rm AB}^{xx} = & \frac{\cal D}{4\rho^3} 
\left(\frac{1}{2}-\frac{3x_{\rm I}y_{\rm I}}{\rho^2} \right) \sin 2\phi_A^0 
\nonumber \\
& \times \left(\phi_{\bf n}^2+\phi_{\bf m}^2 -2\phi_{\bf n}\phi_{\bf m}
+\theta_{\bf n}^2+\theta_{\bf m}^2 \right), \nonumber \\
u_{\rm AB}^{yy} = & \frac{\cal D}{4\rho^3} 
\left(\frac{1}{2}+\frac{3x_{\rm I}y_{\rm I}}{\rho^2} \right) \sin 2\phi_A^0 
\nonumber \\
& \times \left(\phi_{\bf n}^2+\phi_{\bf m}^2 -2\phi_{\bf n}\phi_{\bf m}
+\theta_{\bf n}^2+\theta_{\bf m}^2 \right), \nonumber \\
u_{\rm AB}^{xy} = &\frac{3\cal D}{4\rho^5} \left(x_{\rm I}^2-y_{\rm I}^2\right)  
\nonumber \\
& \times \left(\phi_{\bf n}^2+\phi_{\bf m}^2 +2\phi_{\bf n}\phi_{\bf m}
+\theta_{\bf n}^2+\theta_{\bf m}^2 \right), \nonumber \\
u_{\rm AB}^{zz} = & \frac{\cal D}{\rho^3} \theta_{\bf n}\theta_{\bf m}. 
\end{align}
Then the single AB  pair interaction is
\begin{align}
u_{\rm AB}^{(2)}   = \frac{\cal D}{\rho^3}  \bigl\{
\tfrac{1}{4}\left(\phi_{\bf n}^2+\phi_{\bf m}^2 -2\phi_{\bf n}\phi_{\bf m}
+\theta_{\bf n}^2+\theta_{\bf m}^2 \right) \sin 2\phi_A^0  \nonumber \\
  +\tfrac{3\left(x_{\rm I}^2-y_{\rm I}^2\right)}{4\rho^2} 
\left(\phi_{\bf n}^2+\phi_{\bf m}^2 +2\phi_{\bf n}\phi_{\bf m}
+\theta_{\bf n}^2+\theta_{\bf m}^2 \right) 
+ \theta_{\bf n}\theta_{\bf m} \bigr\}.
\end{align}
This is summed over all AB pairs, which is one contribution to $H^{(2)}$.
Each pair must be summed only once, which can be done by restricting {\bf n} to be
an A-site only and {\bf m} to be a B-site only, denoted as 
\be
U_{\rm AB}^{(2)} = \sum_{\bf n}^{\rm A} \sum_{\bf m}^{\rm B} u_{\rm AB}^{(2)}.
\ee
But the expressions are symmetric in {\bf n} and {\bf m}, so
that summing $\phi_{\bf m}^2$ over {\em all} {\bf m} is equivalent to summing $\phi_{\bf n}^2$
over {\em all} {\bf n}.  That means $\phi_{\bf m}^2$ can be removed, and {\bf n} can be
summed over all sites of both sublattices.  The term $\theta_{\bf m}^2$ can be removed for 
the same reason.  For the cross terms, 
let ${\bf m}={\bf n}+{\bf r}$, sum over the allowed AB displacements {\bf r}, sum
over {\em all} {\bf n}, and then divide by two to undo the double counting of bonds.
{\bf n} might be an A or B site, it doesn't matter, as long as ${\bf r}=(x_{\rm I}, y_{\rm I})$
is an AB bond, which is enforced with $(x_{\rm I}+y_{\rm I})$ being an odd integer.
This gives the AB bond contribution to $H^{(2)}$,
\begin{align}
U_{\rm AB}^{(2)} 
= \sum_{\bf n}\sum_{\bf r}^{\rm AB} \frac{\cal D}{\rho^3}  \Bigl\{
\tfrac{1}{4}\left(\phi_{\bf n}^2-\phi_{\bf n}\phi_{\bf m}
+\theta_{\bf n}^2 \right) \sin 2\phi_A^0  \nonumber \\
  +\tfrac{3\left(x_{\rm I}^2-y_{\rm I}^2\right)}{4\rho^2} 
\left(\phi_{\bf n}^2+\phi_{\bf n}\phi_{\bf m}
+\theta_{\bf n}^2 \right) 
+ \tfrac{1}{2}\theta_{\bf n}\theta_{\bf m} \Bigr\}.
\end{align}

A similar procedure is applied for AA bonds.  Eqs.\ (\ref{AxAx}) through (\ref{AxAy}) give
\begin{align}
u_{\rm AA}^{xx} & =  \frac{\cal D}{2\rho^3} 
\left(\frac{1}{2}-\frac{3x_{\rm I}y_{\rm I}}{\rho^2} \right) \nonumber \\
 \times & \left[(\phi_{\bf n}^2+\phi_{\bf m}^2 +\theta_{\bf n}^2+\theta_{\bf m}^2) \cos^2\!\phi_A^0 
-2  \phi_{\bf n}\phi_{\bf m} \sin^2\!\phi_A^0 \right],
 \nonumber \\
u_{\rm AA}^{yy} & = \frac{\cal D}{2\rho^3} 
\left(\frac{1}{2}+\frac{3x_{\rm I}y_{\rm I}}{\rho^2} \right) \nonumber \\
 \times & \left[(\phi_{\bf n}^2+\phi_{\bf m}^2 +\theta_{\bf n}^2+\theta_{\bf m}^2) \sin^2\!\phi_A^0 
-2  \phi_{\bf n}\phi_{\bf m} \cos^2\!\phi_A^0 \right],
 \nonumber \\
u_{\rm AA}^{xy} & = \frac{3\cal D}{4\rho^5} \left(x_{\rm I}^2-y_{\rm I}^2\right)  
\nonumber \\
& \times \left(\phi_{\bf n}^2+\phi_{\bf m}^2 +2\phi_{\bf n}\phi_{\bf m}
+\theta_{\bf n}^2+\theta_{\bf m}^2 \right) \sin 2\phi_A^0, \nonumber \\
u_{\rm AA}^{zz} & = \frac{\cal D}{\rho^3} \theta_{\bf n}\theta_{\bf m}. 
\end{align}
Their sum is a single AA pair interaction,
\begin{align}
u_{\rm AA}^{(2)} & = \frac{\cal D}{\rho^3} \Bigl\{
\tfrac{1}{4} (\phi_{\bf n}^2+\phi_{\bf m}^2-2\phi_{\bf n}\phi_{\bf m}+\theta_{\bf n}^2+\theta_{\bf m}^2)
+\theta_{\bf n}\theta_{\bf m} 
\nonumber \\
+ & \frac{3(x_{\rm I}^2-y_{\rm I}^2)}{4\rho^2} 
\left(\phi_{\bf n}^2+\phi_{\bf m}^2 +2\phi_{\bf n}\phi_{\bf m}
+\theta_{\bf n}^2+\theta_{\bf m}^2 \right)  \sin 2\phi_A^0  \nonumber \\
- & \frac{3x_{\rm I}y_{\rm I}}{2\rho^2} \left(\phi_{\bf n}^2+\phi_{\bf m}^2 +2\phi_{\bf n}\phi_{\bf m}
+\theta_{\bf n}^2+\theta_{\bf m}^2 \right) \cos 2\phi_A^0  \Bigr\}.
\end{align}
When $u_{\rm AA}^{(2)}$ is summed  over all AA pairs, this gives another contribution to $H^{(2)}$. 
In this case both {\bf n} and {\bf m} must be selected from the A-sites, and the sum is
\be
U_{\rm AA}^{(2)} = \frac{1}{2}\sum_{\bf n}^{\rm A} \sum_{\bf m \ne n}^{\rm A} u_{\rm AA}^{(2)}.
\ee
where $\frac{1}{2}$ undoes the double counting of AA bonds.  But summing $\phi_{\bf m}^2$ 
over all A-sites gives the same as summing $\phi_{\bf n}^2$ over all A-sites.  Therefore this 
can be written indicating that {\bf n} is an A-site while the displacements ${\bf r}=(x_{\rm I},y_{\rm I})$
must be AA bonds, enforced by $\left(x_{\rm I}+y_{\rm I}\right)$ being even integers. This gives 
\begin{align}
\label{UAA2}
U_{\rm AA}^{(2)} = & \sum_{\bf n}^{\rm A} \sum_{\bf r}^{\rm AA} \frac{\cal D}{\rho^3} \Bigl\{
\tfrac{1}{4} (\phi_{\bf n}^2-\phi_{\bf n}\phi_{\bf n+r}+\theta_{\bf n}^2)
+\tfrac{1}{2} \theta_{\bf n}\theta_{\bf m} 
\nonumber \\
+ & \frac{3(x_{\rm I}^2-y_{\rm I}^2)}{4\rho^2} 
\left(\phi_{\bf n}^2+\phi_{\bf n}\phi_{\bf n+r} +\theta_{\bf n}^2 \right)  \sin 2\phi_A^0  \nonumber \\
- & \frac{3x_{\rm I}y_{\rm I}}{2\rho^2}\left(\phi_{\bf n}^2+\phi_{\bf n}\phi_{\bf n+r}
+\theta_{\bf n}^2\right)\cos 2\phi_A^0  \Bigr\}.
\end{align}

Finally there are BB bonds, very similar to AA bonds, however, the terms differ because
the equilibrium dipole directions on the two sublattices are different. Eqs.\ (\ref{BxBx})
through (\ref{BxBy}) give
\begin{align}
u_{\rm BB}^{xx} & =  \frac{\cal D}{2\rho^3} 
\left(\frac{1}{2}-\frac{3x_{\rm I}y_{\rm I}}{\rho^2} \right) \nonumber \\
 \times & \left[(\phi_{\bf n}^2+\phi_{\bf m}^2 +\theta_{\bf n}^2+\theta_{\bf m}^2) \sin^2\!\phi_A^0 
-2  \phi_{\bf n}\phi_{\bf m} \cos^2\!\phi_A^0 \right],
 \nonumber \\
u_{\rm BB}^{yy} & =  \frac{\cal D}{2\rho^3} 
\left(\frac{1}{2}+\frac{3x_{\rm I}y_{\rm I}}{\rho^2} \right) \nonumber \\
 \times & \left[(\phi_{\bf n}^2+\phi_{\bf m}^2 +\theta_{\bf n}^2+\theta_{\bf m}^2) \cos^2\!\phi_A^0 
-2  \phi_{\bf n}\phi_{\bf m} \sin^2\!\phi_A^0 \right],
 \nonumber \\
u_{\rm BB}^{xy} & = \frac{3\cal D}{4\rho^5} \left(x_{\rm I}^2-y_{\rm I}^2\right)  
\nonumber \\
& \times \left(\phi_{\bf n}^2+\phi_{\bf m}^2 +2\phi_{\bf n}\phi_{\bf m}
+\theta_{\bf n}^2+\theta_{\bf m}^2 \right) \sin 2\phi_A^0, \nonumber \\
u_{\rm BB}^{zz} & = \frac{\cal D}{\rho^3} \theta_{\bf n}\theta_{\bf m}.
\end{align}
The displacements that connect BB pairs are the same as for AA pairs, selecting
${\bf r}=(x_{\rm I},y_{\rm I})$ with $\left(x_{\rm I}+y_{\rm I}\right)$ being even 
integers.  Summing appropriately over all the BB pairs gives their contribution to $H^{(2)}$,
\begin{align}
U_{\rm BB}^{(2)} = & \sum_{\bf n}^{\rm B} \sum_{\bf r}^{\rm BB} \frac{\cal D}{\rho^3} \Bigl\{
\tfrac{1}{4} (\phi_{\bf n}^2-\phi_{\bf n}\phi_{\bf n+r}+\theta_{\bf n}^2)
+\tfrac{1}{2} \theta_{\bf n}\theta_{\bf m} 
\nonumber \\
+ & \frac{3(x_{\rm I}^2-y_{\rm I}^2)}{4\rho^2} 
\left(\phi_{\bf n}^2+\phi_{\bf n}\phi_{\bf n+r} +\theta_{\bf n}^2 \right)  \sin 2\phi_A^0  \nonumber \\
+ & \frac{3x_{\rm I}y_{\rm I}}{2\rho^2}\left(\phi_{\bf n}^2+\phi_{\bf n}\phi_{\bf n+r}
+\theta_{\bf n}^2\right) \cos 2\phi_A^0  \Bigr\}.
\end{align}
The term with $x_{\rm I}y_{\rm I}$ is reversed in sign from that in $U_{\rm AA}^{(2)}$, Eq.\ (\ref{UAA2}),
which is obtained more easily by transforming $\phi_{\rm A}^0 \rightarrow \frac{\pi}{2}-\phi_{\rm A}^0$
in going from AA interactions to BB interactions for the same {\bf r}.

The final contribution to $H^{(2)}$ comes from the islands' anisotropies, whose quadratic 
contribution can be obtained from (\ref{UK}),
\be
U_K^{(2)} = \sum_{\bf n} \left[ (K_1 \cos 2\phi_A^0)\phi_{\bf n}^2
+\left(K_1 \cos^2\phi_A^0+K_3\right)\theta_{\bf n}^2 \right].
\ee
Then the total second order Hamiltonian is the sum of all these parts,
\be
\label{H2}
H^{(2)} = U_K^{(2)}+U_{\rm AB}^{(2)}+U_{\rm AA}^{(2)}+U_{\rm BB}^{(2)}.
\ee

\section{Matrix elements of $H^{(2)}$}
The matrix elements needed in the dynamics calculations can be found from the full quadratic
Hamiltonian $H^{(2)}$, using its quadratic form,
\be
H^{(2)} = \tfrac{1}{2} \psi_{\theta}^{\dagger} \bm{M}_{\theta} \psi_{\theta}
+ \tfrac{1}{2} \psi_{\phi}^{\dagger} \bm{M}_{\phi} \psi_{\phi}.
\ee
This is equivalent to a double sum over all {\bf n} and {\bf m},
\be
H^{(2)} = \tfrac{1}{2} \sum_{\bf n,m} \left( M_{\theta,\bf n,m} \theta_{\bf n}\theta_{\bf m}
+  M_{\phi,\bf n,m} \phi_{\bf n}\phi_{\bf m}\right).
\ee
The matrix elements can be found either by inspection of $H^{(2)}$ in (\ref{H2}) or by second
derivatives,
\be
M_{\theta, \bf n,m} = \frac{\partial^2 H^{(2)}}{\partial \theta_{\bf n}\partial \theta_{\bf m}}, \quad
M_{\phi, \bf n,m} = \frac{\partial^2 H^{(2)}}{\partial \phi_{\bf n}\partial \phi_{\bf m}}.
\ee
The factors $\theta_{\bf n}^2$ and $\phi_{\bf n}^2$ appear in all four parts of $H^{(2)}$, so
all {\sc lrd} interactions contribute to on-site ($M_{\bf n,n}$) couplings.  Those matrix elements are
\begin{align}
M_{\theta,\bf n,n} & =  M_{\rm dd} + 2\left(K_1 \cos^2\phi_A^0+K_3\right), \nonumber \\
M_{\phi,\bf n,n} & =  M_{\rm dd} + 2 K_1 \cos 2\phi_A^0, 
\end{align}
where $M_{\rm dd}$ is the {\sc lrd} part, the same for $\theta$ and $\phi$,
\begin{align}
M_{\rm dd}  & = \sum_{\bf r}^{\rm AB} 
\frac{\cal D}{\rho^3}\left[\tfrac{1}{2}\sin 2\phi_A^0+\tfrac{3\left(x_{\rm I}^2-y_{\rm I}^2\right)}{2\rho^2}\right] \\
+ & \sum_{\bf r}^{\rm AA} 
\frac{\cal D}{\rho^3} \left[\tfrac{1}{2}+\tfrac{3\left(x_{\rm I}^2-y_{\rm I}^2\right)}{2\rho^2}\sin 2\phi_A^0
 \mp \tfrac{3x_{\rm I}y_{\rm I}}{2\rho^2} \cos 2\phi_A^0 \right]. \nonumber
\end{align}
For A-sites (B-sites), the last term takes the minus (plus) sign.
%
%
%
%
For an infinite system, however, the sums involving $x_{\rm I}$ and $y_{\rm I}$ are zero due to symmetry.
$M_{\rm dd}$ is the same for A- and B-sites, and depends on sums over $1/\rho^3$ for AB or AA bonds,
\be
M_{\rm dd}= \tfrac{1}{2}{\cal D} \left(s_{\rm AB} \sin 2\phi_A^0 +s_{\rm AA} \right)
\ee
where $s_{\rm AB}$ and $s_{\rm AA}$ were defined in (\ref{sums}).
Note that $-\frac{1}{2}M_{\rm dd}$ already appears in the expression (\ref{H0s}) for 
equilibrium energy per island, $H^{(0)}/N$.

There are also matrix elements connecting different sites, which can be grouped according
to bond type (AB, AA or BB), and depend on the bond displacement ${\bf r}=(x_{\rm I},y_{\rm I})$
or on the distance $\rho=\sqrt{x_{\rm I}^2+y_{\rm I}^2}$.
For the $\theta$ coordinate, they don't depend on the bond type:
\be
M_{\theta, \bf n,n+r}^{\rm AB} = M_{\theta, \bf n,n+r}^{\rm AA} = M_{\theta, \bf n,n+r}^{\rm BB} = \frac{\cal D}{\rho^3}. 
\ee
For the $\phi$ coordinate, the bond type is important:
\begin{align}
M_{\phi, \bf n,n+r}^{\rm AB} & = \tfrac{\cal D}{\rho^3} 
\left[-\tfrac{1}{2}\sin 2\phi_A^0+\tfrac{3\left(x_{\rm I}^2-y_{\rm I}^2\right)}{2\rho^2} \right], \\
M_{\phi, \bf n,n+r}^{\rm AA} & = \tfrac{\cal D}{\rho^3} 
\left[-\tfrac{1}{2}+\tfrac{3\left(x_{\rm I}^2-y_{\rm I}^2\right)}{2\rho^2}\sin 2\phi_A^0
-\tfrac{3x_{\rm I}y_{\rm y}}{\rho^2}\cos 2\phi_A^0 \right]. \nonumber  \\
M_{\phi, \bf n,n+r}^{\rm BB} & = \tfrac{\cal D}{\rho^3} 
\left[-\tfrac{1}{2}+\tfrac{3\left(x_{\rm I}^2-y_{\rm I}^2\right)}{2\rho^2}\sin 2\phi_A^0
+\tfrac{3x_{\rm I}y_{\rm y}}{\rho^2}\cos 2\phi_A^0 \right]. \nonumber 
\end{align}
None of the above matrix elements depend on the site ${\bf n}$, but only on the displacement from {\bf n}
to {\bf n+r}, where ${\bf r}=(x_{\rm I},y_{\rm I})$ in integer island coordinates. 

\subsection{Elements of the dynamic matrices $\bm{m}$ and $\bm{n}$}
\label{elements}
The elements of the dynamic matrices $\bm{m}$ (which acts on a $\theta$ wave function) and $\bm{n}$ (which
acts on a $\phi$ wave function) with all {\sc lrd} interactions are implicitly defined via Eq.\ (\ref{lrdyn}),
with traveling waves inserted.
The $aa$ and $bb$ elements stay within a sublattice, so they involve Fourier sums over $M_{\bf n,n+r}^{\rm AA}$ and
$M_{\bf n,n+r}^{\rm BB}$.  
For the $\theta$ operator $\bm{m}$,
\begin{align}
m_{aa} & = m_{bb}  = \frac{\gamma_{\rm e}}{\mu} \left( M_{\theta,\bf n,n}
+\sum_{\bf r}^{\rm AA} M_{\theta,\bf n,n+r}^{\rm AA}e^{\ii {\bf q}\cdot {\bf r}} \right)
\nonumber \\
&  = \frac{\gamma_{\rm e}}{\mu} \left[ M_{\theta,\bf n,n} + {\cal D} f_{\rm evn}({\bf q})  \right].
\end{align} 
This depends on a Fourier sum over AA displacements,
\be
f_{\rm evn}({\bf q}) \equiv \sum_{\bf r}^{\rm AA} \frac{e^{\ii {\bf q}\cdot {\bf r}}}{\rho^3} 
= \sum_{x_{\rm I}+y_{\rm I}}^{\rm even} 
\frac{\cos[(q_{x_{\rm I}}x_{\rm I}+q_{y_{\rm I}}y_{\rm I})a_{\rm I}]}{(x_{\rm I}^2+y_{\rm I}^2)^{3/2}}.
\ee
The restriction that $x_{\rm I}+y_{\rm I}$ is even keeps the bonds on the same sublattice.
The $ab$ and $ba$ elements are determined by Fourier sums over $M_{\bf n,n+r}^{\rm AB}$,
\begin{align}
m_{ab} = m_{ba} = \frac{\gamma_{\rm e}}{\mu} 
\sum_{\bf r}^{\rm AB} M_{\theta,\bf n,n+r}^{\rm AB}e^{\ii {\bf q}\cdot {\bf r}} 
= \frac{\gamma_{\rm e}}{\mu} {\cal D} f_{\rm odd}({\bf q}),
\end{align}
where the sum is restricted by $x_{\rm I}+y_{\rm I}$ being odd,
\be
f_{\rm odd}({\bf q}) \equiv \sum_{\bf r}^{\rm AB} \frac{e^{\ii {\bf q}\cdot {\bf r}}}{\rho^3}
= \sum_{x_{\rm I}+y_{\rm I}}^{\rm odd}
\frac{\cos[(q_{x_{\rm I}}x_{\rm I}+q_{y_{\rm I}}y_{\rm I})a_{\rm I}]}{(x_{\rm I}^2+y_{\rm I}^2)^{3/2}}.
\ee

\begin{widetext}
For the $\phi$ operator $\bm{n}$, the corresponding matrix elements are
\begin{align}
  n_{aa} & = 
 \frac{\gamma_{\rm e}}{\mu} \left( M_{\phi,\bf n,n}
+\sum_{\bf r}^{\rm AA} M_{\phi,\bf n,n+r}^{\rm AA}e^{\ii {\bf q}\cdot{\bf r}} \right) 
= \frac{\gamma_{\rm e}}{\mu} \left\{  M_{\phi,\bf  n,n} 
+ {\cal D}\left[ \tfrac{3}{2} d_{\rm evn}({\bf q})\sin 2\phi_A^0 
 -3f_{\rm evn}^{xy}({\bf q})\cos 2\phi_A^0 -\tfrac{1}{2}f_{\rm evn}({\bf q})\right]  \right\}.  \nonumber \\
  n_{bb} & = 
 \frac{\gamma_{\rm e}}{\mu} \left( M_{\phi,\bf n,n}
+\sum_{\bf r}^{\rm BB} M_{\phi,\bf n,n+r}^{\rm BB}e^{\ii {\bf q}\cdot{\bf r}} \right) 
= \frac{\gamma_{\rm e}}{\mu} \left\{  M_{\phi,\bf  n,n} 
+ {\cal D}\left[ \tfrac{3}{2} d_{\rm evn}({\bf q})\sin 2\phi_A^0 
 +3f_{\rm evn}^{xy}({\bf q})\cos 2\phi_A^0 -\tfrac{1}{2}f_{\rm evn}({\bf q})\right]  \right\}.  \nonumber \\
n_{ab} & = n_{ba} = \frac{\gamma_{\rm e}}{\mu}
\sum_{\bf r}^{\rm AB} M_{\phi,\bf n,n+r}^{\rm AB}e^{\ii {\bf q}\cdot {\bf r}} 
= \frac{\gamma_{\rm e}}{\mu} {\cal D} \left[ \tfrac{3}{2} d_{\rm odd}({\bf q})-\tfrac{1}{2} f_{\rm odd}({\bf q}) \sin 2\phi_A^0\right].
\end{align}
\end{widetext}
These depend on $f_{\rm e}({\bf q})$ and $f_{\rm o}({\bf q})$ and other Fourier sums, 
\begin{align}
f^{xy}_{\rm evn}({\bf q}) & \equiv \sum_{x_{\rm I}+y_{\rm I}}^{\text{even}} \frac{x_{\rm I}y_{\rm I} 
\cos[(q_{x_{\rm I}} x_{\rm I}+q_{y_{\rm I}} y_{\rm I})a_{\rm I}] }{\left(x_{\rm I}^2+y_{\rm I}^2\right)^{5/2}} ,
\nonumber \\
d_{\rm evn}({\bf q}) & \equiv \sum_{x_{\rm I}+y_{\rm I}}^{\text{even}} \frac{(x_{\rm I}^2-y_{\rm I}^2) 
\cos[(q_{x_{\rm I}} x_{\rm I}+q_{y_{\rm I}} y_{\rm I})a_{\rm I}] }{\left(x_{\rm I}^2+y_{\rm I}^2\right)^{5/2}} ,
\nonumber \\
d_{\rm odd}({\bf q}) & \equiv \sum_{x_{\rm I}+y_{\rm I}}^{\text{odd}} \frac{(x_{\rm I}^2-y_{\rm I}^2) 
\cos[(q_{x_{\rm I}} x_{\rm I}+q_{y_{\rm I}} y_{\rm I})a_{\rm I}] }{\left(x_{\rm I}^2+y_{\rm I}^2\right)^{5/2}} .
\end{align}
%


\section{Eigenvalues of the $4\times 4$ dynamic matrix}
\label{eigen}
The general dynamic equation (\ref{4x4}) has the expanded expression,
\be
\left[\begin{array}{cccc} 0 & 0 & m_{aa} & m_{ab}  \\
  0 & 0 & m_{ba} & m_{bb}  \\
  n_{aa} & n_{ab} & 0 & 0  \\
  n_{ba} & n_{bb} & 0 & 0 \end{array} \right]
\left[ \begin{array}{c} a_{\phi} \\ b_{\phi} \\ \ii a_{\theta} \\ \ii b_{\theta} \end{array} \right]
=  \omega \left[ \begin{array}{c} a_{\phi} \\ b_{\phi} \\ \ii a_{\theta} \\ \ii b_{\theta} \end{array} \right].
\ee
This is $\bm{W}\psi = \omega \psi$,  where $\bm{W}$ is the $4\times 4$ matrix.  The solution requires
the determinant $D(\omega)=\vert \bm{W}-\omega \bm{I}\vert$ to be zero. 
This is
\be
D(\omega) = \left\vert \begin{array}{cccc} -\omega & 0 & m_{aa} & m_{ab} \\ 0 & -\omega & m_{ba} & m_{bb} \\
n_{aa} & n_{ab} & -\omega & 0 \\ n_{ba} & n_{bb} & 0 & -\omega \end{array} \right\vert = 0.
\ee
Evaluating $D(\omega)$ by the first row, the first term is
\begin{align}
D_1(\omega) & = -\omega  \left\vert \begin{array}{ccc}
-\omega & m_{ba} & m_{bb} \\ n_{ab} & -\omega & 0 \\ n_{bb} & 0 & -\omega \end{array} \right\vert
\nonumber \\
& = -\omega \left[-\omega^3+(m_{ba}n_{ab}+m_{bb}n_{bb})\omega\right].
\end{align}
The next term is
\begin{align}
D_2(\omega) & = m_{aa}  \left\vert \begin{array}{ccc}
0 & -\omega &  m_{bb} \\ n_{aa} & n_{ab} &  0 \\ n_{ba} & n_{bb} & -\omega \end{array} \right\vert
\nonumber \\
& = m_{aa} \left[-n_{aa}\omega^2+m_{bb}(n_{aa}n_{bb}-n_{ab}n_{ba}) \right].
\end{align}
The third and last term is
\begin{align}
D_3(\omega) & = -m_{ab}  \left\vert \begin{array}{ccc}
0 & -\omega &  m_{ba} \\ n_{aa} & n_{ab} &  -\omega \\ n_{ba} & n_{bb} & 0 \end{array} \right\vert
\nonumber \\
& = -m_{ab} \left[n_{ba}\omega^2+m_{ba}(n_{aa}n_{bb}-n_{ab}n_{ba}) \right].
\end{align}
The total determinant is the sum, $D(\omega)=D_1+D_2+D_3$, which is quadratic in $\omega^2$, 
\begin{align}
D(\omega) & = \omega^4-\left[m_{aa}n_{aa}+m_{bb}n_{bb}+m_{ab}n_{ba}+m_{ba}n_{ab}\right]\omega^2
\nonumber \\
& +(m_{aa}m_{bb}-m_{ab}m_{ba})(n_{aa}n_{bb}-n_{ab}n_{ba}) = 0.
\end{align}
That is very general, and it is the same as
\be
D(\omega) =  \omega^4-\left(\bm{m}^{T}\cdot \bm{n}\right) \; \omega^2+|\bm{m}| |\bm{n}| = 0.
\ee
That involves a scalar product of the $2\times 2$ matrices and their determinants.
Then the eigenvalues in this rather general case are determined by the quadratic formula,
\be
\label{Awe}
\omega^2 = \frac{1}{2}\left[\left(\bm{m}^{T}\cdot \bm{n}\right)
\pm \sqrt{\left(\bm{m}^{T}\cdot \bm{n}\right)^2-4|\bm{m}| |\bm{n}|}\; \right].
\ee
One can verify that all four eigenvalues are real, and they come in $+/-$ pairs, corresponding to 
opposite directions of propagation. The $\pm$ in the expression gives two fundamental solutions
(higher and lower frequencies), whose dispersion relations are denoted $\omega_{\infty}^{+}({\bf q})$ 
and $\omega_{\infty}^{-}({\bf q})$, where the $\infty$ subscript indicates that all {\sc lrd} interactions are 
included.



\begin{thebibliography}{99}

\bibitem{Skjaervo19}  S. H. Skj\ae rv\o, C.H. Marrows,  R.L. Stamps  and L.J. Heyderman,
        {Nat. Rev. Phys.} \textbf{2}, 13 (2020).

\bibitem{Nisoli13} C. Nisoli, R. Moessner and P. Schiffer, {Rev. Mod. Phys.} \textbf{85} 1473 (2013).

\bibitem{Nguyen17} V.D. Nguyen, Y. Perrin, S. Le Denmat, B. Canals, and N. Rougemaille,
        {Phys. Rev. B} \textbf{96} 014402 (2017).

\bibitem{Morgan11}  J.P. Morgan, A. Stein, S. Langridge and C. Marrows, {Nature Phys.} \textbf{7} 75 (2011).

\bibitem{Silva12} R.C. Silva, F.S. Nascimento, L.A. S. M\'{o}l, W.A. Moura-Melo, and A.R. Pereira,
        New J. Phys. \textbf{14}, 015008 (2012).

\bibitem{Ribeiro+17} I.R.B. Ribeiro, F.S. Nascimento, S.O. Ferreira, W.A. Moura-Melo, C.A.R. Costa, J. Borme, P.P. Freitas,
        G.M. Wysin, C.I.L de Araujo and A.R. Pereira {Scientific Reports} \textbf{7}, 13982 (2017).

\bibitem{Nisoli18} C. Nisoli, in \textit{The Role of Topology in Materials}, pp 85--112,
        S. Gupta and A. Saxena (eds), Springer Series in Solid-State Sciences, vol 189. Springer, Cham. (2018),
        https://doi.org/10.1007/978-3-319-76596-9$_4$

\bibitem{Moller06}  G. M\"{o}ller and R. Moessner Phys. Rev. Lett. \textbf{96}, 237202 (2006).

\bibitem{Moller09} G. M\"{o}ller and R. Moessner, Phys. Rev. B \textbf{80}, 140409(R) (2009).

\bibitem{Mol09} L.A.S. M\'{o}l, R.L. Silva, R.C. Silva, A.R. Pereira, W.A. Moura-Melo,
and B.V. Costa, J. Appl. Phys. \textbf{106}, 063913 (2009).

\bibitem{Wang06} R.F. Wang, C. Nisoli, R.S. Freitas, J. Li, W. McConville, B.J. Cooley,  M.S. Lund, N. Samarth,
        C. Leighton,  V.H. Crespi, and P. Schiffer, {Nature} \textbf{439} 303 (2006).

\bibitem{Ke08} X. Ke, J. Li, C. Nisoli, Paul E. Lammert, W. McConville, R.F. Wang, V.H. Crespi, and P. Schiffer,
        {Phys. Rev. Lett.} \textbf{101} 037205 (2008).

\bibitem{Nisoli10} C. Nisoli, J. Li, X. Ke, D. Garand, P. Schiffer, and V.H. Crespi,
        Phys. Rev. Lett. \textbf{105}, 047205 (2010).

\bibitem{Porro+13} J.M. Porro, A. Bedoya-Pinto, A. Berger, and P. Vavassori, New J. Phys. \textbf{15}, 055012 (2013).

\bibitem{Farhan13} A. Farhan, P.M. Derlet, A. Kleibert, A. Balan, R.V. Chopdekar, M. Wyss, J. Perron, A. Scholl,
        F. Nolting, and L.J. Heyderman, Phys. Rev. Lett. \textbf{111}, 057204 (2013).

\bibitem{Zhang+19} X. Zhang, Y. Lao, J. Sklenar, N.S. Bingham, J.T. Batley, J.D. Watts, C. Nisoli, C. Leighton, and P. Schriffer,
        {APL Materials} \textbf{7} 111112  (2019).
\change{
\bibitem{Kapaklis+12} Vassilios Kapaklis, Unnar B Arnalds, Adam Harman-Clarke, Evangelos Th Papaioannou, Masoud Karimipour,
        Panagiotis Korelis, Andrea Taroni, Peter C W Holdsworth, Steven T Bramwell and Bj\"orgvin Hj\"orvarsson,
        New Journal of Physics {\bf 14}, 035009 (2012).}

\bibitem{Gliga+13} S. Gliga,  A. K\'akay, R. Hertel, and O.G. Heinonen {Phys. Rev. Lett.} \textbf{110} 117205 (2013).

\bibitem{Jung+16} M.B. Jungfleisch, W. Zhang, E. Iacocca, J. Sklenar, J. Ding, W. Jiang, S. Zhang, J.E. Pearson,
        V. Novosad, J.B. Ketterson, O. Heinonen, and A. Hoffmann,
        {Phys. Rev. B} \textbf{93} 100401(R) (2016).

\bibitem{Iacocca+16}  E. Iacocca, S. Gliga, R.L. Stamps, and O. Heinonen, {Phys. Rev. B} \textbf{93} 134420 (2016).

\bibitem{Arroo+19} D.M. Arroo, J.C. Gartside, and W.R. Branford, {Phys. Rev. B} \textbf{100} 214425 (2019).

\bibitem{Lasnier+20} T.D. Lasnier and G.M. Wysin {Phys. Rev. B} \textbf{101} 224428 (2020).

\bibitem{Arora+Das21} N. Arora and P Das, {AIP Advances} \textbf{11} 035030 (2021).

\bibitem{Heisenberg28} W. Heisenberg {Z. Phys.} \textbf{49} 619 (1928).

\bibitem{Wysin+13} G.M. Wysin, W.A. Moura-Melo, L.A.S. M\'ol, and A.R. Pereira {New J. Phys.} \textbf{15} 045029 (2013).

\bibitem{Ising25} E. Ising {Z. Physik} \textbf{31} 253 (1925).

\bibitem{Ostman18} E. \"Ostman, U.B. Arnalds, V. Kapaklis, A. Taroni, and B. Hj\"orvarsson
        {J. Phys.: Condens. Matt.} \textbf{30} 365301 (2018).

\bibitem{Wysin+12} G.M. Wysin, W.A. Moura-Melo, L.A.S. M\'ol, and A.R. Pereira {J. Phys.: Condens. Matter} \textbf{24} 296001 (2012).

\bibitem{Brunn+21} O. Brunn, Y. Perrin, B. Canals, and N. Rougemaille {Phys. Rev. B} \textbf{103} 094405 (2021).

\bibitem{Wysin21} G.M. Wysin, {J. Phys.: Condens. Matter} \textbf{34} 065803 (2021).

\bibitem{Wysin+15} G.M. Wysin, A.R. Pereira, W.A. Moura-Melo, and C.I.L. de Araujo
        {J. Phys.: Condens. Matter} \textbf{27}  076004 (2015).

\bibitem{Rougemaille11} N. Rougemaille, F. Montaigne, B. Canals, A. Duluard, D. Lacour, M. Hehn,
        R. Belkhou, O. Fruchart, S. El Moussaoui, A. Bendounan, and F. Maccherozzi,
        Phys. Rev. Lett. \textbf{106}, 057209 (2011).

\bibitem{Shevchenko17} Y. Shevchenko, A. Makarov, and K. Nefedev {Phys. Lett. A} \textbf{381(5)} 428-434 (2017).

\bibitem{Jiles91} David Jiles, \textit{Introduction to Magnetism and Magnetic Materials}, Ch. 11,
        (London: Chapman and Hall 1991).

\bibitem{Wysin15} G.M. Wysin \textit{Magnetic Excitations \& Geometric Confinement: Theory and Simulations}, Ch. 2,
        (London: IOP Expanding Physics ebook 2015)

\end{thebibliography}
\end{document}